\documentclass[eqsecnum,showpacs,aps,prd,epsf,twocolumn]{revtex4}
\usepackage{amsmath,amssymb}
\usepackage{graphicx}

\newcommand{\gsim}{\mbox{\raisebox{-1.ex}{$\stackrel
      {\textstyle>}{\textstyle\sim}$}}}
\newcommand{\lsim}{\mbox{\raisebox{-1.ex}{$\stackrel
      {\textstyle<}{\textstyle \sim}$}}}
\def\Vec#1{\mbox{\boldmath $#1$}}
\def\til#1{\tilde{#1}} 
\begin{document}
\thispagestyle{empty}
\title {Collision of Domain Walls in asymptotically Anti de Sitter spacetime}
\author{Yu-ichi Takamizu$^{1}$}
\email{takamizu@gravity.phys.waseda.ac.jp}
\author{Kei-ichi Maeda$^{1\,,2\,,3}$}
\email{maeda@gravity.phys.waseda.ac.jp}
\address{\,\\ \,\\
$^{1}$ Department of Physics, Waseda University,
Okubo 3-4-1, Shinjuku, Tokyo 169-8555, Japan\\
$^{2}$ Advanced Research Institute for Science and Engineering,
 Waseda University,
Okubo 3-4-1, Shinjuku, Tokyo 169-8555, Japan\\
$^{3}$ Waseda Institute for Astrophysics, Waseda University,
Okubo 3-4-1, Shinjuku, Tokyo 169-8555, Japan}
\date{\today}

\begin{abstract}
We study collision of two domain walls in 
5-dimensional asymptotically Anti de Sitter spacetime.
This may provide the reheating mechanism of an ekpyrotic (or cyclic) 
brane universe, in which two BPS branes collide and evolve into a hot big
bang universe. We  evaluate a change of scalar field making the domain wall and can investigate the effect of a negative 
cosmological term in the bulk to the collision process and the evolution of our universe. 
\end{abstract}


\maketitle
\section{\label{intro} introduction}
The theory of inflation not only resolves the key theoretical problems of
a big bang theory  such as the flatness and the horizon problems 
\cite{inflation,inflation_review}, 
but also seems to be confirmed by recent observational 
data on CMB \cite{inflation_CMB}. 
However, it is still unclear what is the  inflaton 
in a fundamental unified theory such as 
string/M-theory. 
So far, there has not been found a convincing model, 
although we may have several possible links \cite{KKLMMT, Maeda_Ohta}. 

While, a new paradigm on the early universe, the 
so-called brane world  has been proposed \cite{1, 3}. 
Such speculation has
been inspired by recent developments in 
string/M-theory \cite{Witten, String, Polchinski}.  
There has been tremendous work on this scheme of
dimensional reduction \cite{tremendous_work, 4}, 
where ordinary matter fields are confined on
a lower-dimensional hypersurface, while only gravitational field
propagates throughout all of spacetime. 
Among such  dimensional reduction scenarios, 
Randall and Sundrum (RS)~\cite{4} proposed a new model where
four-dimensional Newtonian gravity is recovered at low energies
even without compact extra dimensions. 
Based on such a new world picture, 
many cosmological scenarios have been studied 
\cite{brane_cosmology,
brane_cosmo2,Lukas}. 
See also recent reviews \cite{Maeda,Maartens,Langlois,Brax}.
We have found some 
deviations from standard cosmology by modifications of 
4-dimensional Einstein equations on the brane \cite{shiromizu}, even 
for the case that there is a scalar field in bulk \cite{Maeda_Wands}. 

\ In such a  brane world scenario, for resolving 
the above-mentioned key theoretical problems
in the big bang theory, a 
new idea of cosmological model has been proposed, which is called 
the ekpyrotic scenario or the cyclic universe scenario 
\cite{Khoury, Khoury_cyclic}. 
It could  be an alternative to 
an inflationary scenario. 
Since it is not only motivated by such fundamental unified 
theory but also may resolve the above mentioned key cosmological problems, 
it would be very attractive. 

This scenario is based on a collision of two cold branes. 
The universe starts with a cold, empty, and 
nearly BPS ground state, which contains two parallel branes at rest. 
The two branes approach each other and then collide.
The energy is 
dissipated on the brane and the big bang universe starts. 
The BPS state is required in order to retain a supersymmetry
in a low-energy 4-dimensional 
effective action. The visible and hidden 
branes are flat and are described by a Minkowski spacetime,
 but the bulk is warped along the fifth dimension. 

There has been much discussion about 
density perturbations to see 
whether this scenario is really a reliable scenario for the early 
universe \cite{density_ekpyrotic, density_ekpyrotic2,
density_ekpyrotic3, density_ekpyrotic4}. 
It has been shown that the initial spectrum 
seems not to be 
produced as a scale invariant one.
So this is the weakness of 
this scenario.  
However, it may turn out that 
we find other wayout to produce a scale invariant fluctuation.
Then, it is necessary to analyze the other important process
in this scenario, e.g. the 
reheating process,  in detail.  
For the work about the collision of branes,  there has been some studies by 
\cite{LMW, kofman} 
with using the brane approximated as 
the delta function. 
In the analysis using a delta function, 
the microscopic processes in collision such as
dissipation of matter field and reheating of the universe
 can not be described. 
Even though there are some works on \cite{Barnaby}, 
the reheating mechanism itself 
has not been so far investigated in detail. 
So the purpose of our work is 
to construct a consistent brane collision model and 
to analyze its reheating process. 

With such analysis, we may find how we can recover the hot big bang universe 
after the collision  of branes. 
In our previous work (paper I) \cite{Takamizu:2004rq}, 
we proposed a new reheating mechanism in the ekpyrotic universe, 
such as a quantum creation of particles confined on
the brane   at the collision of two branes. 
As a thick brane, we could
adopt a domain wall model constructed by a 
bulk scalar  field  and analyze the collision of two moving domain walls 
in a 5-dimensional bulk spacetime.
In the paper I,  we consider the simplest situation 
such that two domain walls collide 
in a given background spacetime, i.e. the
five-dimensional (5D) Minkowski spacetime. 
As shown by several authors 
\cite{fractal, 13, 14, 15}, 
 the results highly depend on the incident velocity $\upsilon$ of the walls. 
We have shown the time evolution of colliding two 
domain walls in the paper I, which 
confirms the previous works. For a sufficiently large velocity, e.g. 
$\upsilon\gsim 0.25$, it is shown that a kink and antikink will just bounce off once, because there is no time to exchange the energy 
during the collision. For a lower velocity, we find multiple bounces 
when they collide.  For example, 
we find the bounce 
occurs once for $\upsilon=0.4$, while it does twice for $\upsilon=0.2$. We 
also find many bounce solutions for other incident velocities, as shown in 
Appendix B of \cite{Takamizu:2004rq} (see also \cite{fractal}). 
The number of the bounces sensitively depends on the 
incident velocity. A set of the values 
of $\upsilon$ which give the same number of bounce forms a fractal structure 
in the $\upsilon$-space as shown Fig. 6 of \cite{fractal}. 
Then, we  evaluate a production rate of particles confined to
the domain wall. As a result, 
the energy density of created particles is given as 
$\rho \approx  20 \bar{g}^4 N_b ~m_\Phi^4 $ where $\bar{g}$ is a coupling
constant of particles to a domain-wall scalar field, $N_b
$ is the number of bounces at  the collision and $m_\Phi$ is a fundamental
mass scale of the domain wall. It does not depend on the width  $d$ of the
domain wall, although the typical energy scale of  created
particles is given by $\omega\sim 1/d$. 
The  reheating temperature is evaluated as
$T_{\rm R}\approx 0.88 ~ \bar{g} ~ N_b^{1/4}$.
In order to have the baryogenesis  at the electro-weak energy scale,
the fundamental mass scale is constrained
as $m_\eta \gsim 1.1\times 10^7$ GeV for $\bar{g}\sim 10^{-5}$.
For such an energy scale, we may find a sufficiently hot
universe after collision of domain walls.

In order to study whether such a reheating process is 
still efficient in
more reliable cosmological models, we have to include
the curvature effect. In particular, some brane universe
are discussed with a negative cosmological constant \cite{4}
Hence, we  study here how gravitational effects change
our previous results. 
In order to investigate such an effect, 
we have not only to 
investigate the collision of domain walls in a curved spacetime,
but also to solve the spacetime by use of the 5D Einstein equations. 
Inspired by the RS brane model, 
we include a potential of the scalar field
which provides an effective negative cosmological constant 
in a bulk spacetime. 

We first set up  the initially moving two domain walls,
each of which is obtained by boosting an
exact static domain wall solution \cite{sakai} (\S II). 
Although this solution is obtained in the four dimensions, 
it is easy to extend it to the 5D one.
We then solve the 5D Einstein equations and the dynamical equation 
for a scalar field to analyze  collision of thick walls in 
asymptotically AdS spacetime (\S III). 
The concluding remarks follow in \S IV.

We use the unit of $c=\hbar=1$.

%%%%%%%%%%%%%%%%%%%%%%%%%%%%%%%%%%
%%%%%%%%%%%%%%%%%%%%%%%%%%%%%%%%%%
\section{\label{brane colli}
collision of two domain walls} 
%%%%%%%%%%%%%%%%%%%%%%%%%%%%%%%%%
%%%%%%%%%%%%%%%%%%%%%%%%%%%%%%%%%%
\subsection{Basic Equations}
%%%%%%%%%%%%%%%%%%%%%%%%%%%%%%%%%%
%%%%%%%%%%%%%%%%%%%%%%%%%%%%%%%%%%
We study 
collision of two domain walls in 5D
spacetime.
To construct a domain wall structure, we adopt
a 5D real scalar field $\Phi$ 
with an appropriate potential $V(\Phi)$,
which minimum value is  negative. 
This potential gives an asymptotically anti-de Sitter (AdS)
spacetime just as the RS brane model.

Since we discuss the collision of two parallel domain walls, the scalar
field is assumed to depend only on a time coordinate $t$ and  one spatial
coordinate $z$.
 The remaining three spatial
coordinates are denoted by $\Vec{x}$. 
For numerical analysis, we use dimensionless parameters and
variables, which are rescaled  by the mass scale $m_{\Phi}$,
which is  defined by
the vacuum expectation value 
at a local minimum as $\Phi_0=m_\Phi^{3/2}$, as
\begin{equation}
\ \til{t}=m_\Phi t \,,
\ \til{z}=m_\Phi z \,,
\til{\Phi}=\frac{\Phi}{m_\Phi^{3/2}}\,.
\end{equation}
In what follows, we drop the tilde in dimensionless variables for brevity.

It is possible to choose coordinates such that the bulk metric has the 
``2D conformal gauge" , i.e.
\begin{equation}
ds^2=e^{2A(t,z)}(\,-dt^2+dz^2\,)+e^{2B(t,z)}d\Vec{x}^2\,.
\label{z-metric}
\end{equation}
This gauge choice also makes the initial setting easy
 when we construct  moving domain walls
by use of the Lorentz boost. 

In this gauge, the 5D Einstein equations 
and the dynamical equation for a scalar field 
 are split into 
three dynamical equations; 
\begin{align}
\ddot{A}=&A''+3\dot{B}^2-3{B'}^2
-\kappa_5^2(\dot{\Phi}^2-{\Phi'}^2+\frac{1}{3}e^{2A}V(\Phi))\,,\nonumber\\
\ddot{B}=&B''-3\dot{B}^2+3{B'}^2+\frac{2}{3}\kappa_5^2e^{2A}V(\Phi)\,,\nonumber
\\
\ddot{\Phi}=&\Phi''-3\dot{B}\dot{\Phi}+3B'\Phi'-\frac{1}{2}e^{2A}V'(\Phi)\,,
\label{Dynamical equation}
\end{align}
plus two constraint equations; 
\begin{align}
&\dot{B}B'-A'\dot{B}-\dot{A}B'+\dot{B}'=
-\frac{2}{3}\kappa_5^2\dot{\Phi}\Phi'\,,\nonumber\\
&2{B'}^2+B''-A'B'-\dot{A}\dot{B}-\dot{B}^2\nonumber\\ 
&=-\frac{1}{3}\kappa_5^2
(\dot{\Phi}^2+{\Phi'}^2+e^{2A}V(\Phi))\,,
\label{Constraint equation}
\end{align}
where a dot ($\dot{}$) and a prime ($'$)
  denote ${\partial}/{\partial t}$ and 
 ${\partial}/{\partial z}$, respectively. 

These are our basic equations.
Before solve them numerically, we have to set up
our initial data, which satisfies the constraint equations
(\ref{Constraint equation}).

\subsection{Domain Wall Solution}
\begin{figure}[htbp]
\begin{center}
\includegraphics[width=4cm]{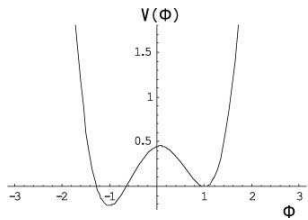}
\caption{
The scalar field potential $V(\Phi)$ is plotted where $d=\sqrt{2}, 
\kappa_5=0.3$. 
This potential behaves like a double-well potential for $|\Phi|< 5$, 
except that one of its two minima has a negative value. On the other hand, 
for $|\Phi|> 5$,  
this potential behaves differently from a double-well potential, that is 
$V(\Phi)$ rapidly falls into $-\infty$.}
\label{fig1}
\end{center}
\end{figure}

For an initial configuration of a domain wall, 
we use an exact static solution given by \cite{sakai}.
They assume a scalar field  $\Phi$ with a potential  
\begin{equation}
V(\Phi)=\Bigl(\frac{\partial W}{\partial \Phi}\Bigr)^2
-\frac{8}{3}\kappa_5^2\,W^2\,,
\end{equation}
where 
\begin{equation}
W \equiv {1\over d}\left(\Phi-\frac{1}{3}\Phi^3-{2\over 3}
\right)
\end{equation}
is a superpotential, and $\kappa_5^2$ and $d$ are the 
five dimensional gravitational constant and 
the thickness of a domain wall, respectively.
The potential minima are located at $\Phi=\pm 1$ 
 in the range of $|\Phi|\lsim 5$ 
 (see Fig. \ref{fig1}). 
The potential 
shape is similar to
a double-well 
potential, but it is asymmetric.
The minimum value at $\Phi=1$ vanishes, while
that at $\Phi=-1$ is negative.
With this potential, we can obtain analytically 
an exact solution 
for two colliding domain walls as follows.
%%%%%%%%%%%%%%%%%%%%%
%%%%%%%%%%%%%%%%%%%%%%%%%%%%%%%%

First,  we show a static domain wall solution with this 
potential.
A kink solution of a scalar field ($K$) is 
described as 
\begin{equation}
\Phi_K(y)=\tanh\left(\ \frac{y}{d}\ \right)\,,
\label{kink}
\end{equation}
and the metric of 5D spacetime is 
\begin{equation}
ds^2=e^{2A_K(y)}(-dt^2+d\Vec{x}^2)+dy^2\,,
\label{y-metric}
\end{equation}
with 
\begin{equation}
A_K(y)=-\frac{4}{9}\kappa_5^2
\Bigl\{\ln \left[\cosh\left({y\over d}\right)\right]
+ \frac{\tanh^2 ( y/d )}{4}-{y\over d}\Bigr\}\,.
\label{asym_AdS_metric}
\end{equation}
Since this exact solution is not given in our gauge,
we have used a new coordinate $y$, which will be 
associated with $z$ later.

\begin{figure}[htbp]
\begin{center}
\includegraphics[width=6cm]{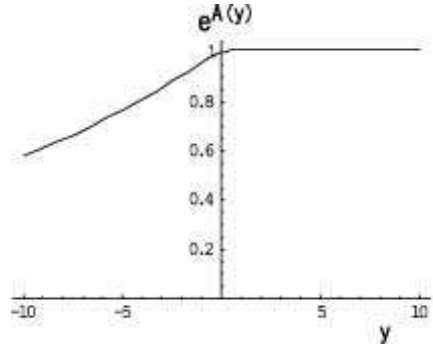}
\caption{The metric component of the
exact solution for a static domain wall \cite{sakai} 
is plotted where we set $d=\sqrt{2}$ and $
\kappa_5=0.3$. This spacetime approaches 
Minkowski space because $A_K \rightarrow A_\infty$ (a constant), 
while it becomes asymptotically AdS because $A_K\rightarrow
-k|y|$ as $y\to -\infty$, where $k=8\kappa_5^2/9d$. }
\label{fig2}
\end{center}
\end{figure}

This metric approaches that of the AdS spacetime in
one asymptotic region ($y\ll -1$), i.e. 
\begin{equation}
e^{2A_K}\rightarrow e^{-2k|y|}\, ~~~{\rm as}~~ y\to -\infty\,,
\end{equation}
with
\begin{equation}
k=\frac{8\kappa_5^2}{9d}\,.
\end{equation}
While it becomes a flat Minkowski space in another asymptotic region
($y \gg 1$), i.e.
 \begin{equation}
e^{2A_K}\rightarrow e^{2A_\infty}\, ~~~{\rm as}~~ y\to \infty\,,
\end{equation}
with
\begin{equation}
A_\infty=\frac{4}{9}\kappa_5^2\left(\log 2-\frac{1}{4}\right)\,.
\label{A_infty}
\end{equation}
We depict the behaviour of metric function $\exp[A_K(y)]$ in
 Fig. \ref{fig2}.

%%%%%%%%%%%%%%%%%%%%%
%%%%%%%%%%%%%%%%%%%%%%%%%%%%%%%%%

By reflecting the 
spatial coordinate $y$,
we also find an antikink solution
($\bar{K}$)  as
$\Phi_{\bar{K}}(y)=\Phi_K(-y)=-\Phi_K(y)$.  
The corresponding metric  of this antikink solution 
is also obtained by reflection of $y$-coordinate, i.e. 
$A_{\bar{K}}(y)=A_K(-y)=-A_K(y)$. 

In order to describe this solution under our gauge
condition (\ref{z-metric}), that is,
 in the $(t,z)$ frame, 
we should transform the present ($t,y$) coordinates  (Eq.(\ref{y-metric})) 
into the ($t,z$) ones (Eq.(\ref{z-metric})) by defining  
the coordinate $z$ as
\begin{equation}
z=\int e^{-A_K(y)}dy  \,.
\label{translate}
\end{equation}
This integration will be performed numerically 
to find initial data of  collision of two domain walls.

\subsection{Moving Domain Wall}

When a domain wall moves with constant speed $\upsilon$ in the fifth 
direction $z$, 
we can obtain the corresponding solution 
by boosting a static kink solution ($K$) as 
\begin{equation}
\Phi_\upsilon(z,t)=\tanh\left[\ \frac{1}{d}~
y^*\left(\gamma(z-\upsilon
t)\right)\
\right]
\,,
\label{moving_DW}
\end{equation} 
where $y^*$ and $z^*$ are comoving coordinates 
of a domain wall, and
$y^*(z^*)$
is  
obtained by the inverse transformation of 
Eq. (\ref{translate}). 
The Lorentz 
transformation gives 
$z^*=\gamma(z-\upsilon
t)$ where
$\gamma=1/\sqrt{1-\upsilon^2}$\ is the Lorentz factor. 
We have assumed that the center of
a domain wall is initially located at $z=0$.

The corresponding metric is easily obtained 
by Lorentz boost.
Because of the  Lorentz invariance in our 2D conformal gauge, i.e., 
${-d{t^*}^2+d{z^*}^2}=-dt^2+dz^2$, 
we find
\begin{equation}
ds^2_{2D}=\exp [{2A_K\left(\gamma(z-\upsilon t)\right)}](-dt^2+dz^2)\,,
\label{moving_DW2}
\end{equation}
where $A_K(z^*)=A_K (y^*(z^*))$. 
The  function $A_K(y)$ is given by Eq. (\ref{asym_AdS_metric}).
The center of  a domain wall ($z^*=0$) moves as 
$z=\upsilon t$ in our $(t,z)$-coordinate  frame.
Then we regard that the metric describes a spacetime with a
domain wall moving 
with constant speed $\upsilon$ in the $z$ direction as well as a 
scalar field $\Phi$
does so.

In order to discuss collision of two moving domain walls, 
we first have to set up its initial data.
Using Eqs. (\ref{moving_DW}) and (\ref{moving_DW2}), 
we can construct an initial data for two moving
domain walls
as follows. 
Provide a kink solution at $z=-z_0$ and an antikink solution 
at $z=z_0$, which are separated by a 
large distance and approaching each other with the same speed 
$\upsilon$. 
We then obtain the following initial data;
\begin{align}
&\Phi(z,0)=
\Phi_\upsilon(z+z_0,0)-\Phi_{-\upsilon}(z-z_0,0)-1 \,,\nonumber\\
&A(z,0)=
A_\upsilon(z+z_0,0)-A_{-\upsilon}(z-z_0,0)-A_\infty \,,
\label{initialPhi} 
\end{align}
where $A_\infty$ is the constant value given by Eq. (\ref{A_infty}). 
 The initial values of $\dot{\Phi}$ and $\dot{A}$ are also given by 
\begin{align}
&\dot{\Phi}(z,0)=
\dot{\Phi}_\upsilon(z+z_0,0)-\dot{\Phi}_{-\upsilon}(z-z_0,0)\,,\nonumber\\
&\dot{A}(z,0)=
\dot{A}_\upsilon(z+z_0,0)-\dot{A}_{-\upsilon}(z-z_0,0)\,.
\label{initialPhidot} 
\end{align}
Obviously, we set $A=B$ and $\dot{A}=\dot{B}$ at initial. 

The  spatial separation between two  walls is given by $2z_0$, and 
as long as the separation distance is much larger
than the thickness of the wall ($z_0\gg d$),
the initial conditions (\ref{initialPhi}) 
and (\ref{initialPhidot}) give a good 
approximation for two moving domain walls. 
Using these initial values, we  solve
the dynamical  equation (\ref{Dynamical equation}) 
numerically. The results will be
shown in the next section.

%%%%%%%%%%%%%%%%%%%%%%%%%%%%%%%%%%%
%%%%%%%%%%%%%%%%%%%%%%%%%%%%%%%%%%%%
\section{Time Evolution of Colliding Domain Walls}
%%%%%%%%%%%%%%%%%%%%%%%%%%%%%%%%%%
%%%%%%%%%%%%%%%%%%%%%%%%%%%%%%%%%%

We use a numerical approach to solve 
the dynamical equations (\ref{Dynamical equation}) 
for the colliding domain walls. 
We adopt a numerical method similar to 
one used in \cite{Takamizu:2004rq}. 
The difference is found in a boundary conditions.
We impose the Dirichlet boundary condition for the scalar 
field, which is the same as the paper I, 
while the Neumann boundary condition 
is used for
 the metric as $A'(z)=-k\gamma e^{A(z)}$, which is derived from 
the asymptotic form of the metric, i.e., $e^{A(z)}\to 1/(k(\gamma |z|+1))$ 
as $|z|\to \infty$. 

We have three free parameters in our model, i.e. 
a wall thickness 
$d$, an initial wall velocity
${\upsilon}$, and a warp factor ${k}$ (or 
the gravitational constant $\kappa_5$). 
Two of them ($d, k ( {\rm or}~ \kappa_5)$) are
 fundamental because they appear in the 
theory.
In the paper I, 
we studied the collision of two domain 
walls in the fixed Minkowski background \cite{fractal},
where we had two free parameters
$d$ and $\upsilon$.
So, including the gravitational back reaction, we
investigate how $\kappa_5$ (or  $k$) changes the previous results. 
In what follows, fixing the value of $d$, i.e. $d=\sqrt{2}$,
we show our results.

%%%%%%%%%%%%%%%%%%%%%%%%%%%%%%%%%%%%%%%%%%%%
%%%%%%%%%%%%%%%%%%%%%%%%%%%%%%%%%%%%%%%%%%%%%
\subsection{Time Evolution of Scalar Field $\Phi$}
%%%%%%%%%%%%%%%%%%%%%%%%%%%%%%%%%%%%%%%%%%%%%
%%%%%%%%%%%%%%%%%%%%%%%%%%%%%%%%%%%%%%%%%%%%%%

First let us set $\kappa_5=0$ (or $k=0$), 
 that is the case of Minkowski background spacetime. 
Although the scalar field potential is slightly different 
from that in the paper I, 
the result is exactly the same. 
This simulation also gives a check of our numerical code. 

\begin{figure}[ht]
\begin{center}
\includegraphics[width=8cm]{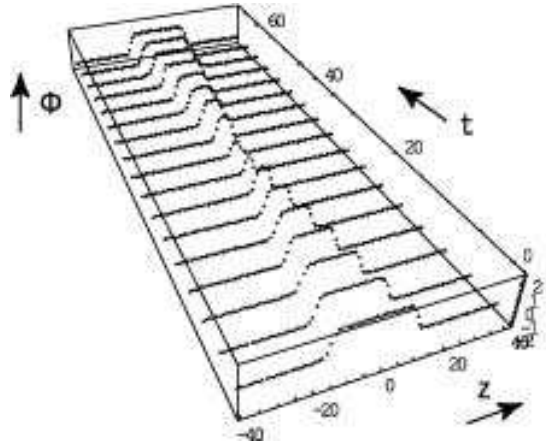}
\caption{Collision of two domain walls 
where the initial velocity $\upsilon=0.4$. 
The time evolution of the scalar field $\Phi$ is shown from $t=0$ to 70.
The collision occurs once around $t=31$.
We set $d=\sqrt{2}$, $\kappa_5=0.05$. This process is 
similar to the Minkowski case, $\kappa_5=0$. } 
\label{fig3}
\end{center}
\end{figure}

\begin{figure}[ht]
\begin{center}
\includegraphics[width=8cm]{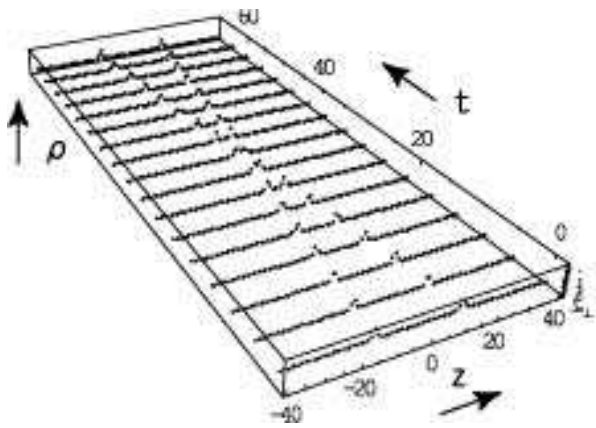}
\caption{The time evolution of scalar field 
energy density in  the
case of Fig. \ref{fig3}.
The maximum point of $\rho_{\Phi}$ defines the position
of a wall ($z=z_{\rm W}(t)$).} 
\label{fig4}
\end{center}
\end{figure}

Next we perform our simulation
for $\kappa_5\neq 0$.
For a small value of $\kappa_5$, i.e., $\kappa_5\lsim 0.05$ 
(equivalently 
 $k\lsim 1.57 \times10^{-3}$ 
or
$m_\Phi \lsim (0.05)^{2/3} m_5\approx 0.136 m_5 $),
the collision process is very similar to the case 
of the Minkowski background. 
Setting  the initial velocity  $\upsilon =0.4$,
we show our numerical results for $\kappa_5=0.05$
in Figs. \ref{fig3} and 
\ref{fig4}. 
The evolution of $\Phi$
 is depicted in Fig. \ref{fig3}, 
while that of the energy density is shown in Fig. \ref{fig4}. 
%%%%%%%%%%%%%%%%%%%%%
%%%%%%%%%%%%%%%%%%%%%%%%%%%%%%%%%
%%%%%%%%%%%%%%%%%%%%%
%%%%%%%%%%%%%%%%%%%%%%%%%%%%%%%%%
%%%%%%%%%%%%%%%%%%%%%
\begin{figure}[htbp]
\begin{center}
\includegraphics[width=8cm]{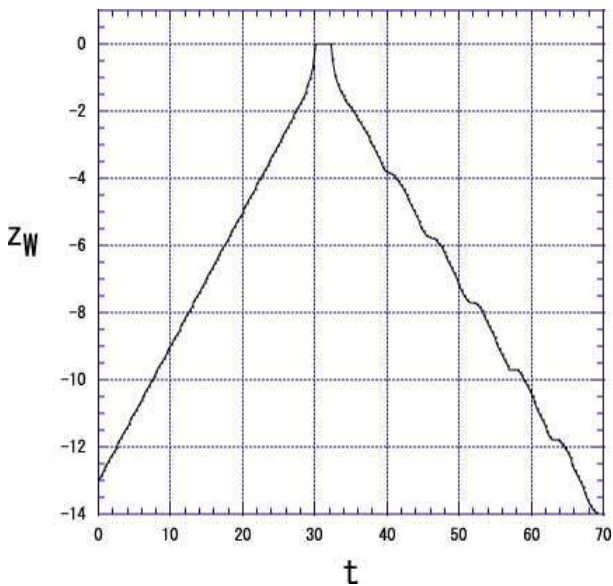}
\caption{Time evolution of the position of one brane $z=z_{\rm W}$ which 
starts to move at $z=-13$  with a constant speed $\upsilon=0.4$. 
We set $d=\sqrt{2},\ \kappa_5=0.05$. } 
\label{fig position}
\end{center}
\end{figure}
%%%%%%%%%%%%%%%%%%%%%%%%%%%%%%%%%

The energy density of the scalar field 
is given by 
\begin{equation}
\rho_{\Phi}=e^{-2A}\Bigl(\dot{\Phi}^2+{\Phi'}^2\Bigr)
+V(\Phi)\,.
\end{equation}
We find some peaks in the energy density, by which  we
 define the positions of moving walls ($z=\pm z_{\rm W}(t)$).
If a domain wall is symmetric, its position is defined  by 
$\Phi(z)=0$. However, in more general situation, just as in the present case
that the scalar field is oscillating around some moving point, it
may be natural  to define the position of a domain wall by the maximum
point of its energy density. 

Fig. \ref{fig position} denotes the position of brane 
$z=z_{\rm W}(t)$ with respect to $t$. 
The brane moves with constant 
speed $\upsilon=0.4$ toward the collision point $z=0$,
and collide, 
then recede to the boundary. 
We also find  small oscillation 
around a uniform motion after collision.

Since we assume that we are living on one domain wall, we are interested 
 in a particle  $\psi$ confined on the  domain wall.
If a particle $\psi$ is coupled with a 5-D scalar field 
$\Phi$, which is responsible for the domain wall, we expect
quantum production of $\psi$-particles at collision of domain walls.
This is because the value of the scalar field $\Phi$ 
on the domain wall changes with time. 
This fact may play an 
important role in a reheating mechanism \cite{Takamizu:2004rq}. 
Once we find the solution of colliding domain walls, we know
the time evolution of a scalar field on the domain wall,
and we can evaluate production rate. 

At the position of a domain wall,
 the induced metric is given as 
\begin{equation}
ds^2=-d\tau^2+a^2(\tau)d\Vec{x}^2\,,
\end{equation}
where proper time is determined as 
\begin{equation}
\tau=\int e^{A_{\rm W}}dt\,,
\label{proper time}
\end{equation}
and the scale factor is 
\begin{equation}
a(\tau)=e^{B_{\rm W}}(\tau)\,.
\label{scale factor}
\end{equation}
$A_{\rm W}$ and  $B_{\rm W}$ are evaluated on the brane, 
i.e., $A_{\rm W}=A(t, z_{\rm W}(t))$ and $B_{\rm W}=B(t, z_{\rm W}(t))$. 
The Hubble parameter of the brane universe
is defined by
\begin{equation}
H(\tau)\equiv \frac{1}{a}\frac{da}{d\tau}\Big|_{z=z_{\rm W}}=
\frac{dB_{\rm W}}{d\tau}=
e^{-A_{\rm W}}\dot{B}_{\rm W}(\tau)\,. 
\label{hubble}
\end{equation}
 
%%%%%%%%%%%%%%%%%%%%%
\begin{figure}[htbp]
\begin{center}
\includegraphics[width=8cm]{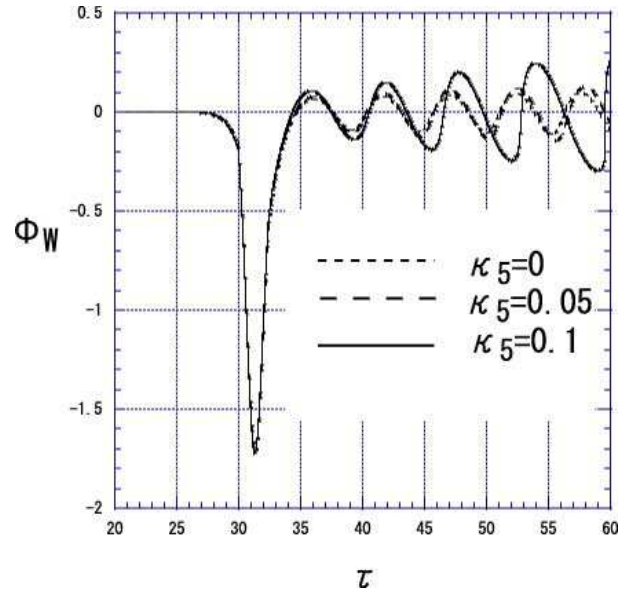}
\caption{Time evolution of a scalar field 
on the moving wall 
with $\upsilon=0.4,\ d=\sqrt{2}$. 
We set $\kappa_5=0,0.05, 0.1$.
The value of the scalar field on the wall
is defined by  
$\Phi_{\rm W}=\Phi(t, z_{\rm W}(t))$, where $z_{\rm W}(t)$
 is the position of the wall. The time of period of
 oscillation gets slightly longer and 
the amplitude is a little bit larger as $\kappa_5$ increases.} 
\label{fig5}
\end{center}
\end{figure}
\begin{figure}[ht]
\begin{center}
\includegraphics[width=8cm]{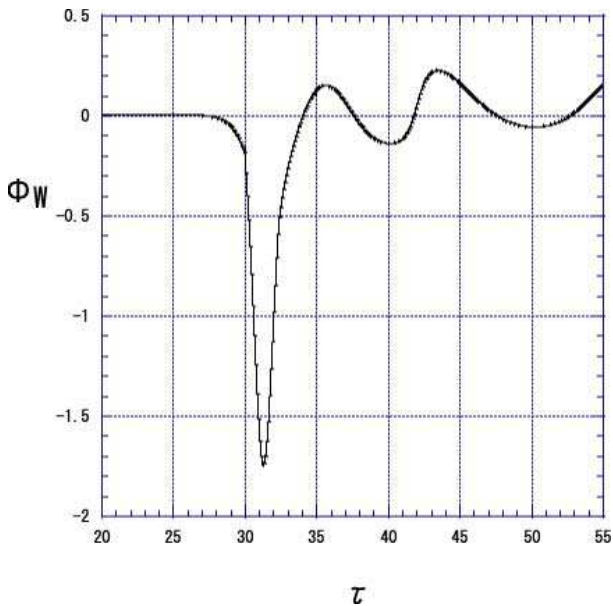}
\caption{Time evolution of a scalar field 
on the moving wall 
for $\upsilon=0.4,\ d=\sqrt{2},\ \kappa_5=0.15$. 
The value of the scalar field is given by  
$\Phi_{\rm W}=\Phi(t, z_{\rm W}(t))$, where $z_{\rm W}(t)$ is the 
position of the wall. $\Phi_{\rm W}$ goes out of oscillation phase 
$0$ at $\tau\sim55$ and 
then numerical simulation stops.} 
\label{fig8}
\end{center}
\end{figure}

%%%%%%%%%%%%%%%%%%%%%%%%
For $\upsilon=0.4$, we depict the time evolution of 
$\Phi_{\rm W}$ on one moving wall 
for different values of $\kappa_5$ 
in Figs.
\ref{fig5} and \ref{fig8}. 
The feature of collision is similar,
 but the behaviour of a scalar field on 
the moving wall after collision is different for each $\kappa_5$. 

We summarize our numerical results 
for each value of $\kappa_5$ in order.
\\[.5em]
(i) $\kappa_5$=0.01\\[.5em]
For a small value of $\kappa_5$, e.g., 
$\kappa_5=0.01$ (equivalently $k\sim 6.29 \times 10^{-5}$ 
or $m_\Phi \sim 0.0464 m_5 $), 
the result is almost the same as the case of 
the Minkowski background, in which case 
we find one bounce point, 
which corresponds to a crossing point
 in Fig. \ref{fig4}, and then
the oscillations around $\Phi_W=0$ follow
(see the dotted line in Fig. \ref{fig5}). 
This oscillation 
is explained by using a perturbation analysis 
in Minkowski spacetime \cite{Takamizu:2004rq}. 
We have found one stable oscillation mode around the kink solution.
This oscillation appears by excitation of the system at collision. 
\\[.5em]
(ii) $\kappa_5=0.05$\\[.5em]
As increasing the value of  $\kappa_5$ 
 slightly larger,
 for example $\kappa_5=0.05$, (equivalently for 
$k\sim 1.57 \times 10^{-3}$   or $m_\Phi
\sim 0.136 m_5$), 
the time evolution of $\Phi_{\rm W}$ slightly 
changes. 
In Fig. \ref{fig5}, just as 
the case of $\kappa_5=0$,
we find one bounce and successive oscillations. 
However, the period of oscillation is slightly
longer and the amplitude 
gets a little bit larger as $\kappa_5$ increases. 
\\[.5em]
(iii) $\kappa_5= 0.15$\\[.5em]
For $\kappa_5= 0.15$, (equivalently $k\sim 1.41 \times 10^{-2}$ 
or  $m_\Phi
\sim 0.282 m_5$), 
the behaviour of this oscillation changes drastically. 
After several oscillations, the scalar field leaves $\Phi_W=0$ 
as shown in Fig. \ref{fig8}. The numerical simulation 
eventually breaks down because all variables diverge.

%%%%%%%%%%%%%%%%%%%%%%%%%%%%%%%%%
%%%%%%%%%%%%%%%%%%%%%
%%%%%%%%%%%%%%%%%%%%%
%%%%%%%%%%%%%%%%%%%%%%%%%%%%%%%%%%%
\begin{figure}[htbp]
\begin{center}
\begin{tabular}{cc}
\includegraphics[width=4cm]{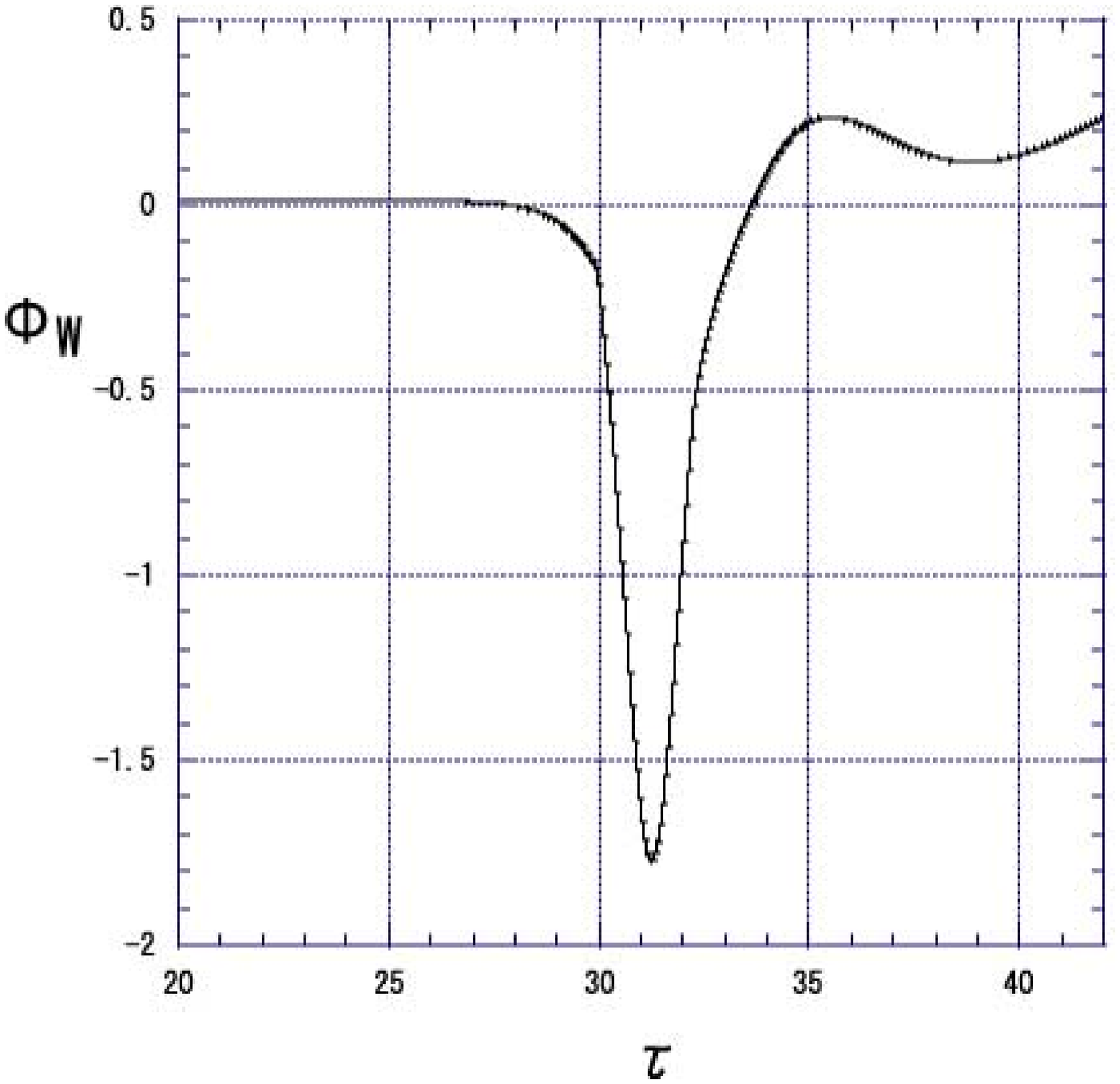} &
\includegraphics[width=4cm]{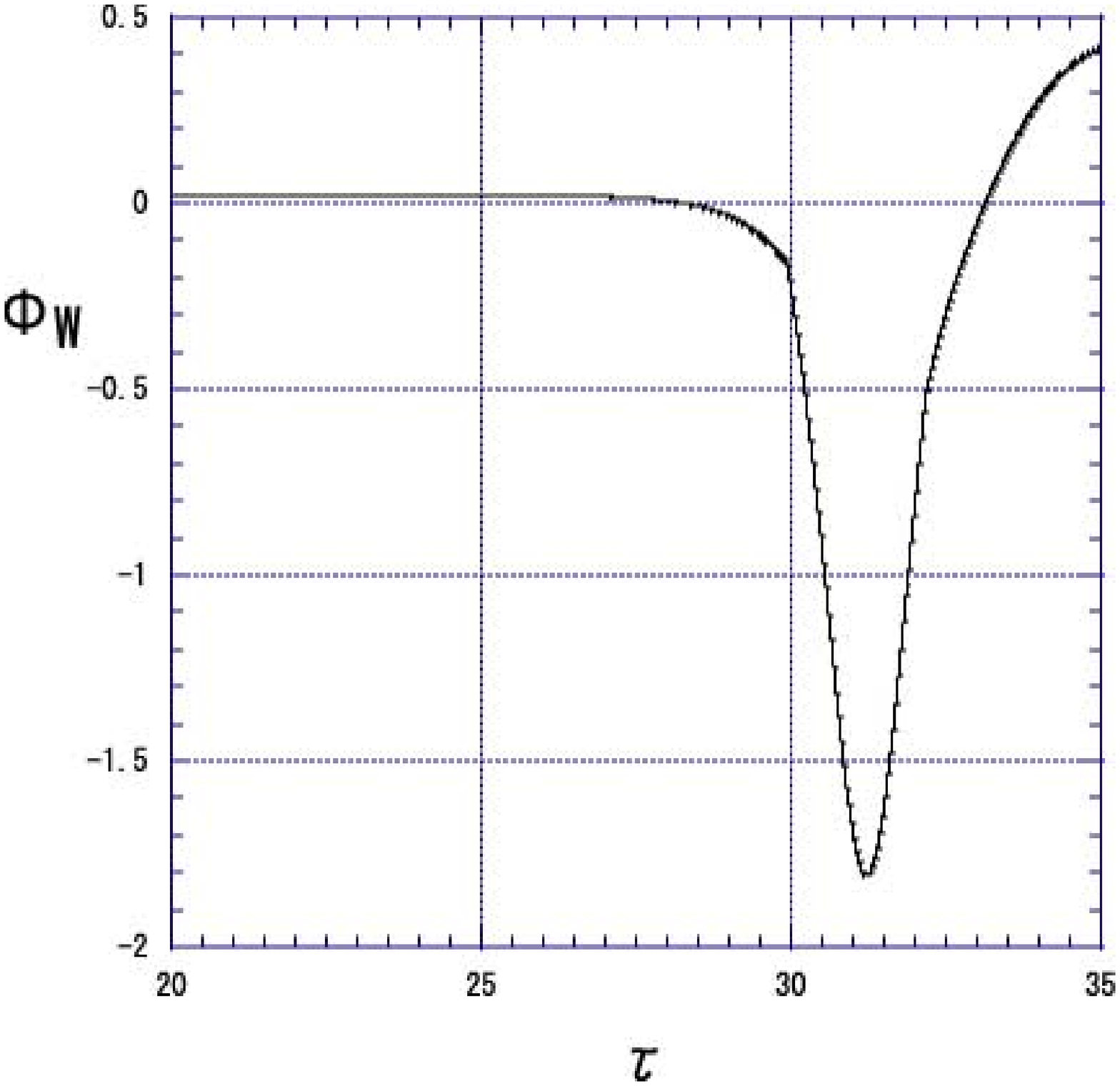}\\
(a) $\kappa_5=0.2$ & (b) $\kappa_5=0.25$ \\
\end{tabular}
\caption{Time evolution of the scalar field $\Phi_{\rm W}$ 
for $\upsilon=0.4,\ d=\sqrt{2}$,
setting (a) $\kappa_5=0.2$  and (b) $\kappa_5=0.25$. 
The numerical simulation breaks down 
when (a) $\tau\sim 42$ and  (b) $\tau\sim 35$, 
respectively.}
\label{fig13}
\end{center}
\end{figure}
%%%%%%%%%%%%%%%%%%%%%%%%%%%%%%%%

(iv)  $\kappa_5 > 0.15$\\[.5em]
For the larger values of $\kappa_5$ than 0.15, 
the time to appearance of singularity
 becomes shorter, that is, contrary to  the case of 
$\kappa_5\lsim 0.15$, the scalar field after 
collision does not oscillate but leave $\Phi_{\rm W}=0$ soon. 
The time evolution of  $\Phi_{\rm W}$ is shown in Fig. \ref{fig13}, 
for $\kappa_5=0.2$ ($k=2.51 \times 10^{-2}$) and 
$\kappa_5=0.25$ ($k=3.93 \times 10^{-2}$). 

\begin{figure}[htbp]
\begin{center}
\begin{tabular}{cc}
\includegraphics[width=4cm]{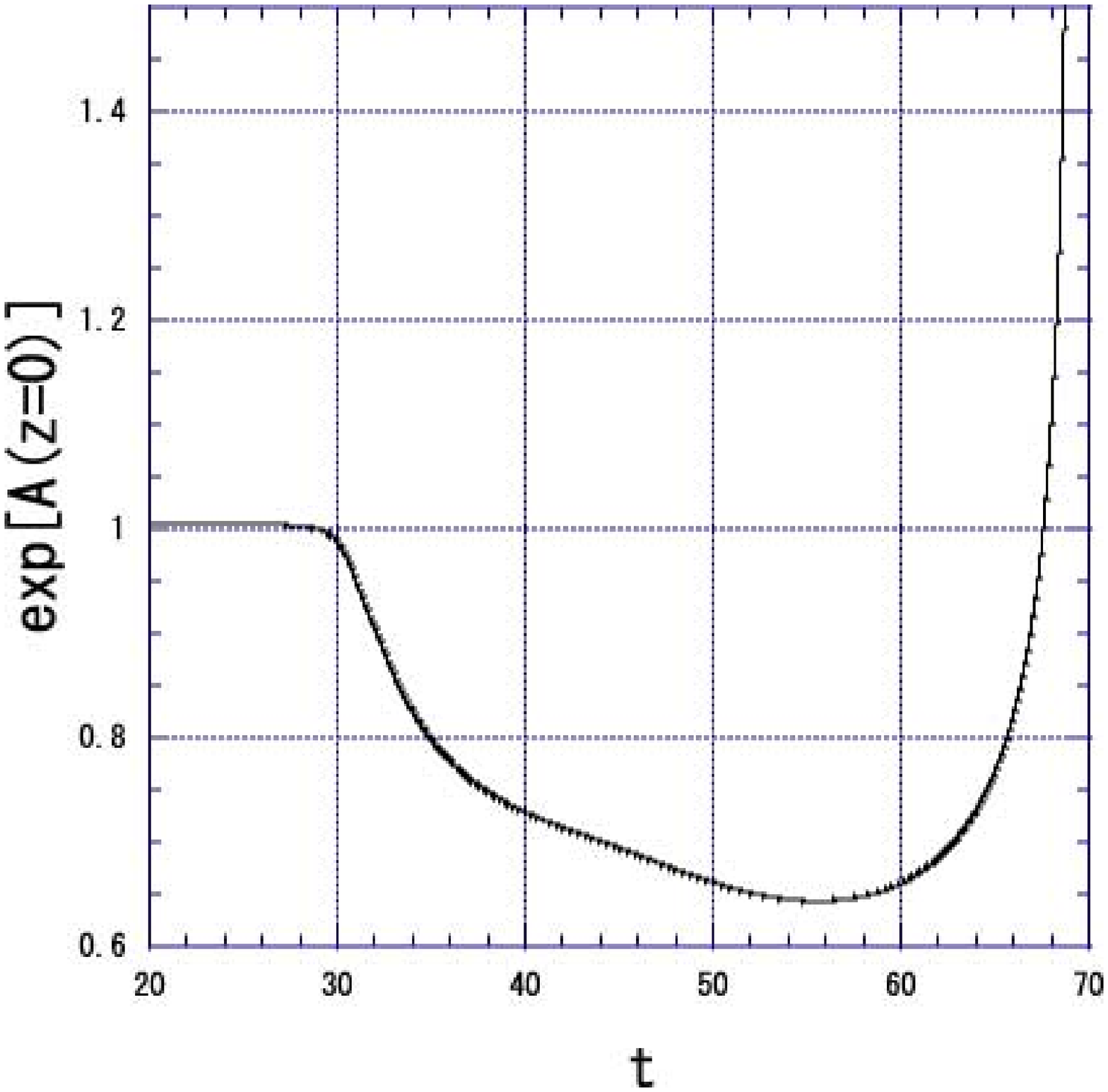} &
\includegraphics[width=4cm]{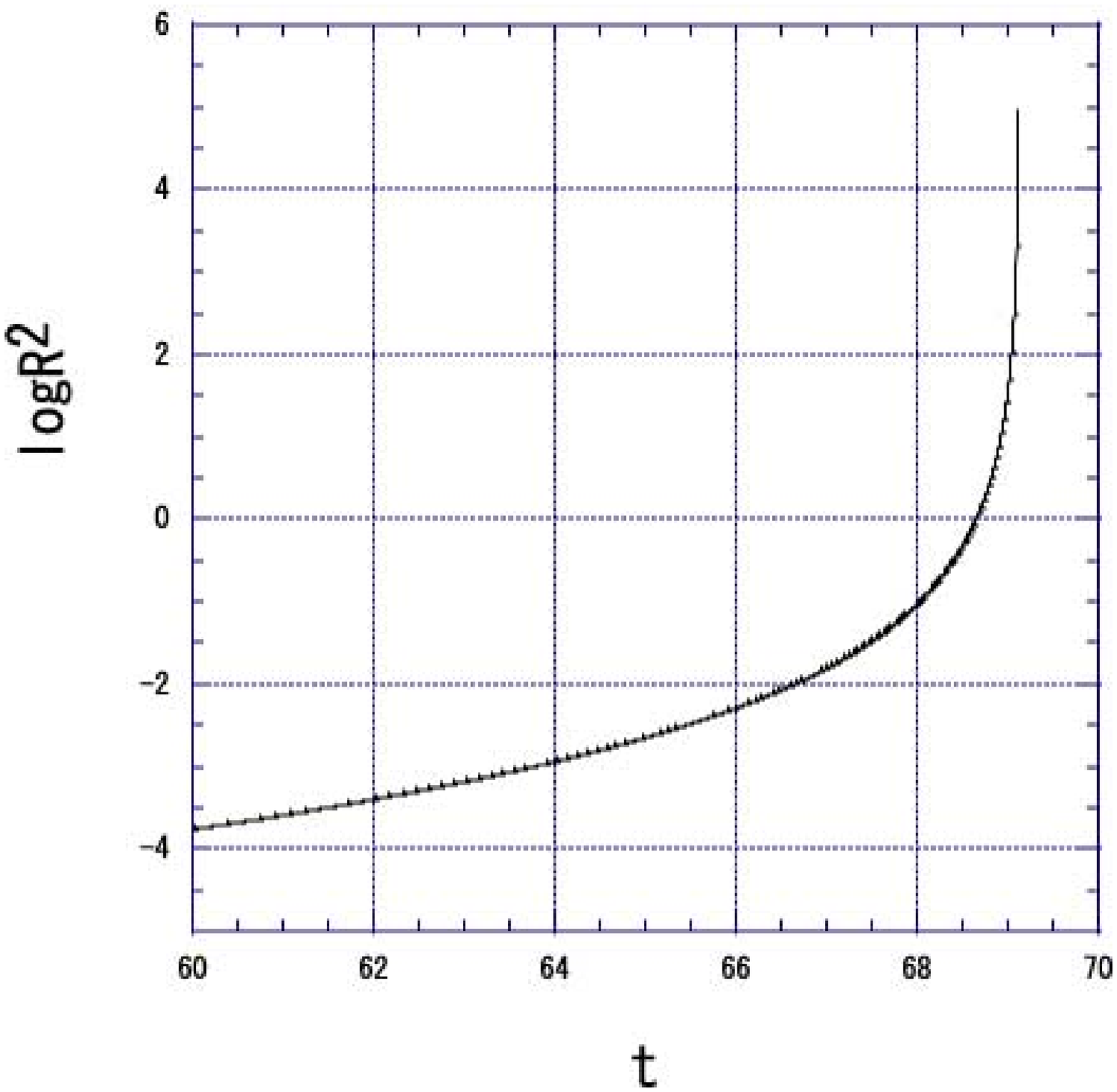}\\
(a) & (b) \\
\end{tabular}
\caption{ (a) Time evolution of the metric component $A$ at $z=0$ 
for $\upsilon=0.4,\ d=\sqrt{2},\ \kappa_5=0.15$. 
It diverges at $t\sim 70$. 
(b)  Kretschmann scalar invariant ($R^{abcd}R_{abcd}$) at 
the origin $z=0$ 
for the case of $\upsilon=0.4, d=\sqrt{2}, \kappa_5=0.15$. 
We find that the Kretschmann scalar diverges at $t\simeq 69$.
This means that it is not a coordinate singularity, but a
curvature singularity.
}
\label{fig12}
\end{center}
\end{figure}

The metric component $A$ at $z=0$ also diverges as shown
in Fig \ref{fig12}. 
It is not 
 a coordinate singularity, but a curvature
singularity.
In order to show it,
we calculate the so-called Kretschmann invariant
scalar, which is 
the simplest scalar invariant quadratic in the Riemann tensor, 
and is defined as
\begin{align}
&R^{abcd}R_{abcd}=
e^{-4A}\Bigl[3(\ddot{B}+\dot{B}^2-\dot{A}\dot{B}-A'B')^2
\nonumber\\
&
-3(A'\dot{B}+\dot{A}B'-\dot{B}'-\dot{B}B')^2
 +(\ddot{A}-A'')^2
\nonumber\\ 
&+3({B'}^2-\dot{B}^2)^2+3(B''+{B'}^2-\dot{A}\dot{B}-A'B')^2
\Bigr]\,.
\end{align}

In Fig. \ref{fig12}(b), we depict the time evolution of
 the Kretschmann scalar at the origin $z=0$, which 
diverges at $t\simeq 69$. 
It is caused by the divergence of a quantity $\dot{A}$. 
We conclude a singularity forms at the origin $z=0$. 

This divergence is not a numerical error because a constraints equations 
(\ref{Constraint equation}) are always satisfied
within $10^{-5}$ - $10^{-2}$ \% accuracy except
at time when the singularity appears.

In Appendix \ref{appen:perturb}, we study 
the reason why a spacetime is unstable and eventually
evolves into a singularity in detail 
using a perturbation analysis of the Einstein equations and
dynamical equation of $\Phi$. We find that
a perturbed oscillation mode 
 around an unperturbed
kink solution becomes overstable 
for $\kappa_5=0.1$.
From our analysis, we conclude that
 gravitational back reaction makes
a kink solution unstable 
contrary to the Minkowski case. 

~\\[-0.5em]
%%%%%%%%%%%%%%%%%%%%%%%%%%%%%%%%%
%%%%%%%%%%%%%%%%%%%%%
%%%%%%%%%%%%%%%%%%%%%%%%%%%%%%%%%%%
\begin{figure}[htbp]
\begin{center}
\begin{tabular}{cc}
\includegraphics[width=4cm]{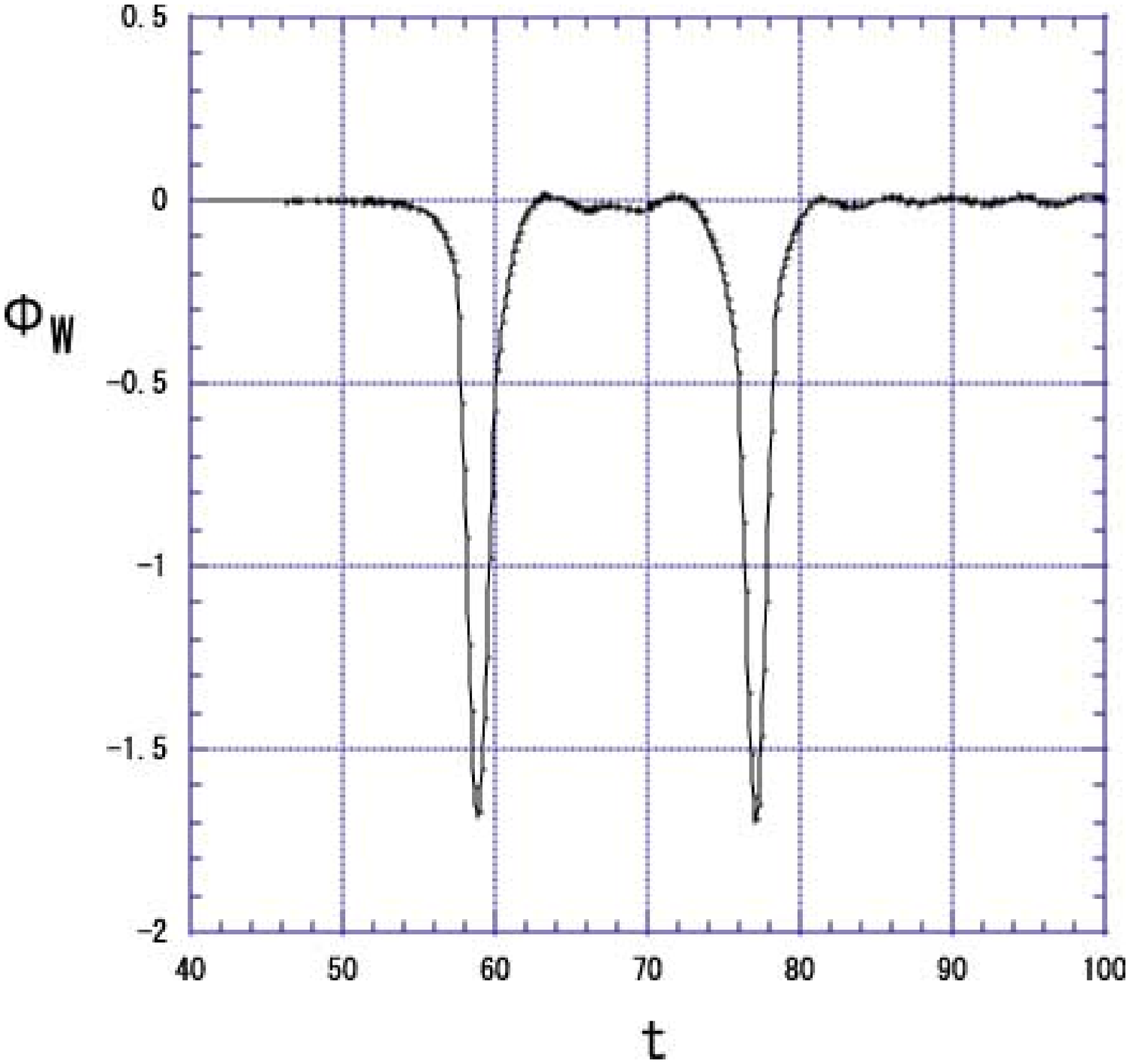} &
\includegraphics[width=4cm]{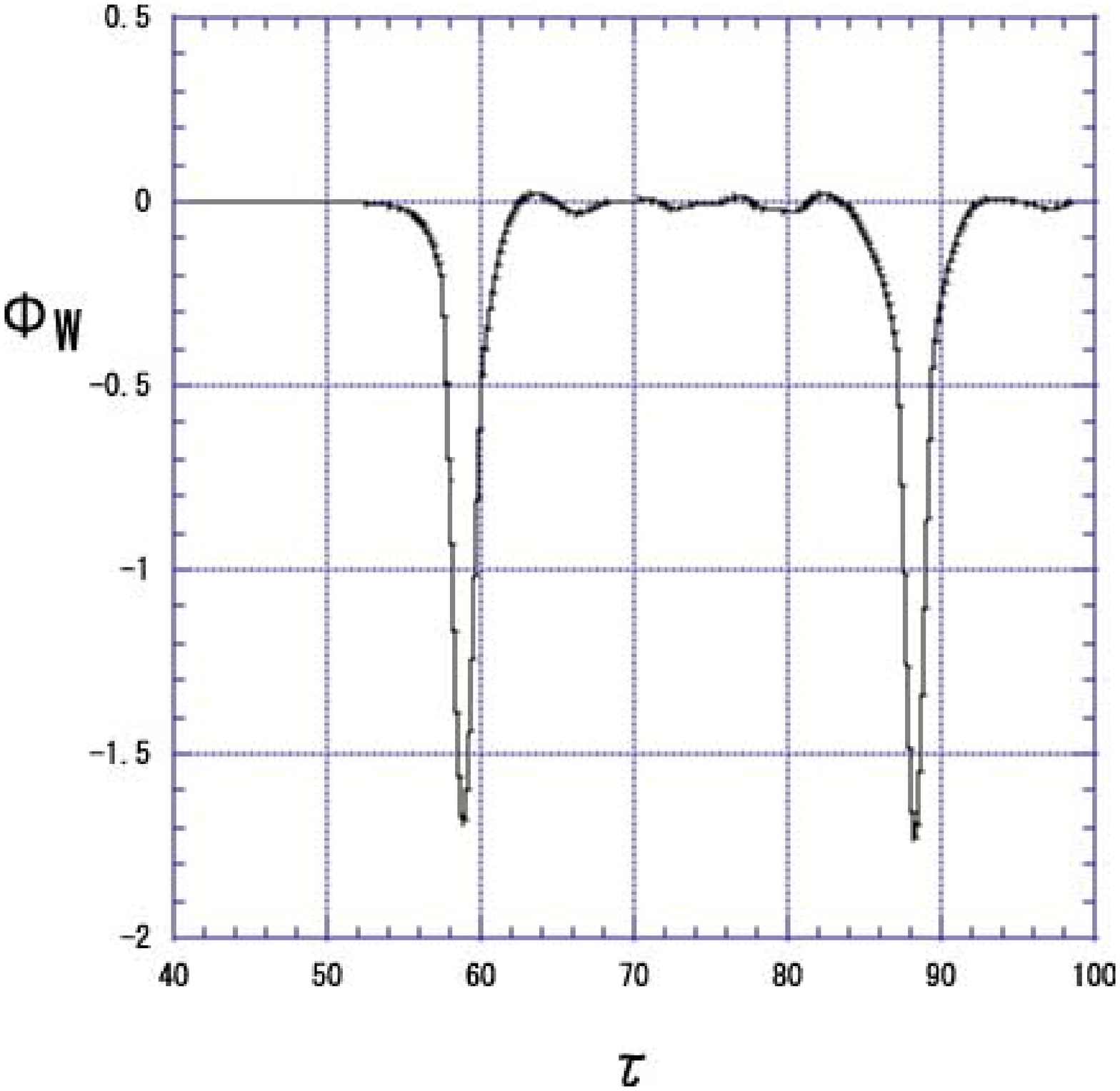}\\
(a) $\kappa_5=0$ (Minkowski) & (b) $\kappa_5=0.05$ \\
\end{tabular}
\caption{Time evolution of a scalar field $\Phi_{\rm W}$ 
for $\upsilon=0.2,\ d=\sqrt{2}$.
For (a)
$\kappa_5=0$ and (b) $\kappa_5=0.05$, we find two peaks 
which correspond to 
twice bounces at collision. Moreover, it is seen that
an effective
negative cosmological constant prolongs the time interval  between 
two bounces.}
\label{v=0.2_phi}
\end{center}
\end{figure}
%%%%%%%%%%%%%%%%%%%%%%%%%%%%%%%%
%%%%%%%%%%%%%%%%%%%%%%%%%%%%%%
\begin{figure}[htbp]
\begin{center}
\includegraphics[width=8cm]{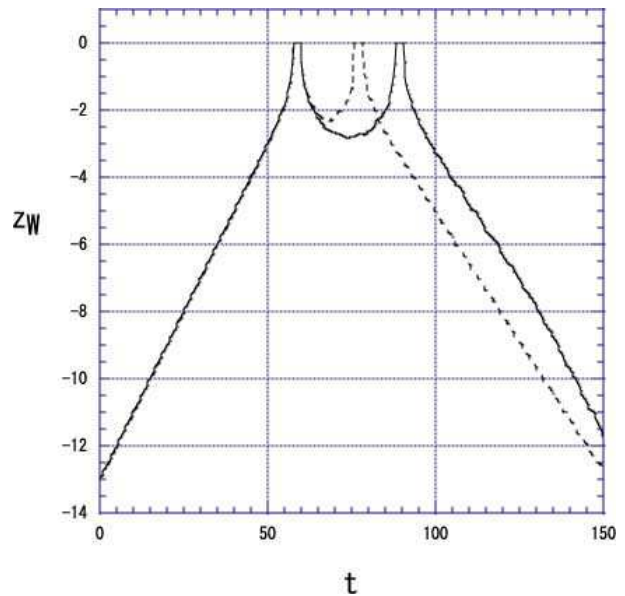}
\caption{Time evolution of the position of one brane $z=z_{\rm W}$ which 
starts to move at  $z=-13$  with a constant speed $\upsilon=0.2$, setting
$d=\sqrt{2} \ \kappa_5=0.05$. The dashed line denotes $\kappa_5=0.05$ while 
the dotted line does $\kappa_5=0$.
The time interval between 
two bounces is prolonged by 
an effective  negative cosmological constant. } 
\label{v=0.2_position}
\end{center}
\end{figure}
%%%%%%%%%%%%%%%%%%%%%%%%%%%%%%%%%%%
%%%%%%%%%%%%%%%%%%%%%%%%%%%%%%
\begin{figure}[htbp]
\begin{center}
\includegraphics[width=8cm]{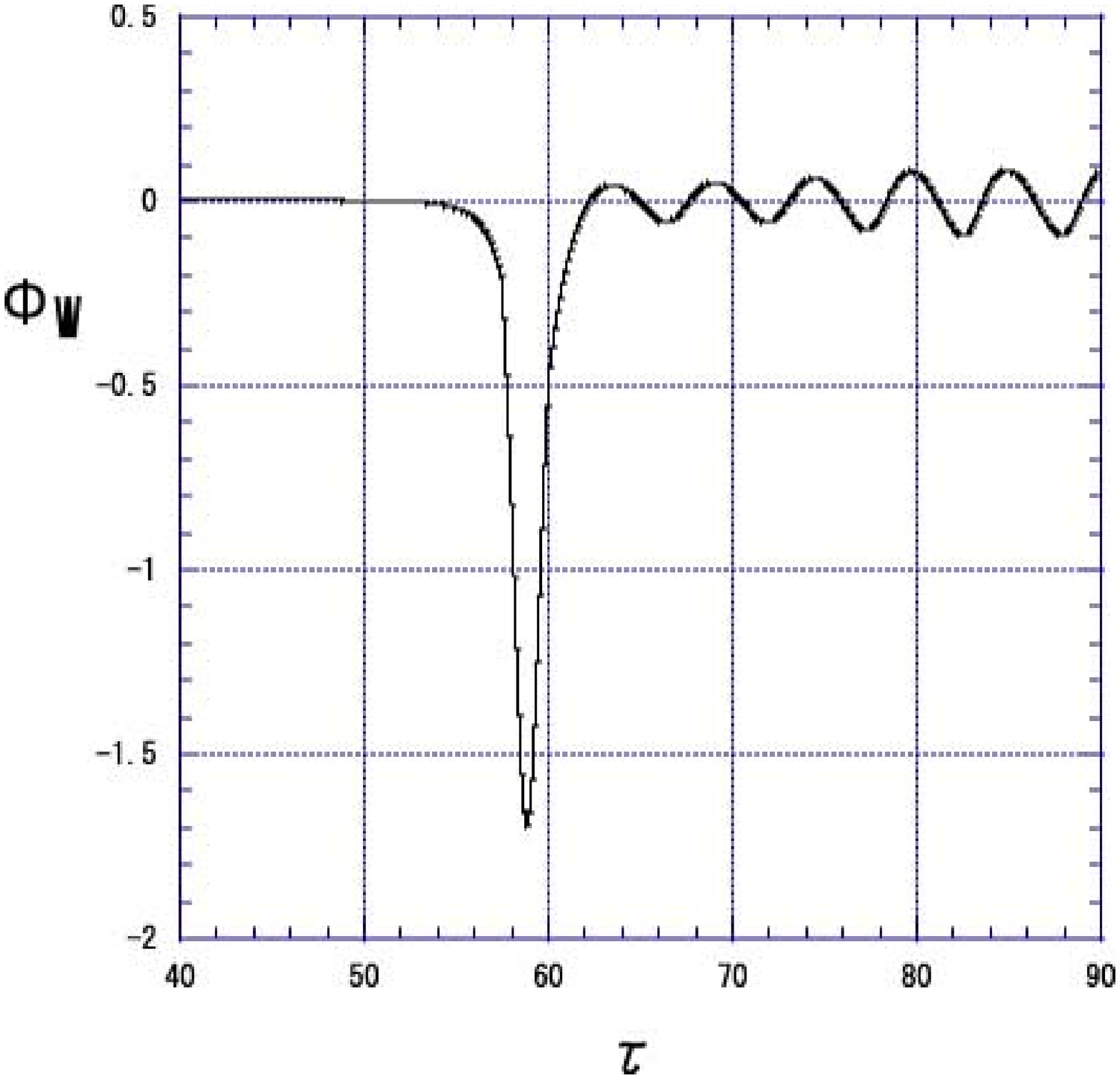}
\caption{Time evolution of the scalar field $\Phi_{\rm W}$ 
for $\upsilon=0.2$,  \ $d=\sqrt{2},  \ \kappa_5=0.1$. 
Two-bounce process does not occur. 
Compare it with the cases in Fig. \ref{v=0.2_phi}. } 
\label{v=0.2_phi2}
\end{center}
\end{figure}
%%%%%%%%%%%%%%%%%%%%%%%%%%%%%%%%%%%

Next, we show the result for the case of 
the initial velocity $\upsilon=0.2$. In the Minkowski case, 
this incident velocity shows two-bounce at collision process
(see Fig. \ref{v=0.2_phi} (a)).  
After two walls collide, they bounce, recede to 
some finite distance, turn back and then collide again. 
For small 
values of $\kappa_5$, 
e.g., $\kappa_5\leq 0.05$, the collision process is very similar
to the case of
 $\kappa_5=0$
(see Fig. \ref{v=0.2_phi}(b)). 
As $\kappa_5$ increases,  the time interval between 
first and second bounces becomes longer
as
shown in Fig. \ref{v=0.2_position}. 
This can be understood from the fact  that 
the above mentioned oscillation after collision will
radiate the energy.
So a kink-antikink pair is loosely bounded and
it takes longer time to collide again.

For the case of $\kappa_5\gsim 0.1$, this feature of collision 
is drastically changed. 
Two-bounce collision never occurs, that is, two 
walls collide only once as shown in 
Fig. \ref{v=0.2_phi2}. This is because a lot of energy of a
kink-antikink pair is radiated away via the unstable oscillation 
after collision and  it has not enough 
energy to form a trapped state.
After the first bounce, the domain walls never collide again but recede 
each other.

For larger value of $\kappa_5$,
we find only one-bounce collision. 
Namely, a ``negative cosmological constant"
outside a kink-antikink pair 
keeps away two walls toward the boundary, 
so it plays as an effective attractive force. 

%%%%%%%%%%%%%%%%%%%%%%%%%%%%%%%%%%%%%%%%%%%%
%%%%%%%%%%%%%%%%%%%%%%%%%%%%%%%%%%%%%%%%%%%%%
\subsection{Time Evolution of Metric}
%%%%%%%%%%%%%%%%%%%%%%%%%%%%%%%%%%%%%%%%%%%%%
%%%%%%%%%%%%%%%%%%%%%%%%%%%%%%%%%%%%%%%%%%%%%%

%%%%%%%%%%%%%%%%%%%%%%%%%%%%%%%%%%%
\begin{figure}[htbp]
\begin{center}
\begin{tabular}{cc}
\includegraphics[width=4cm]{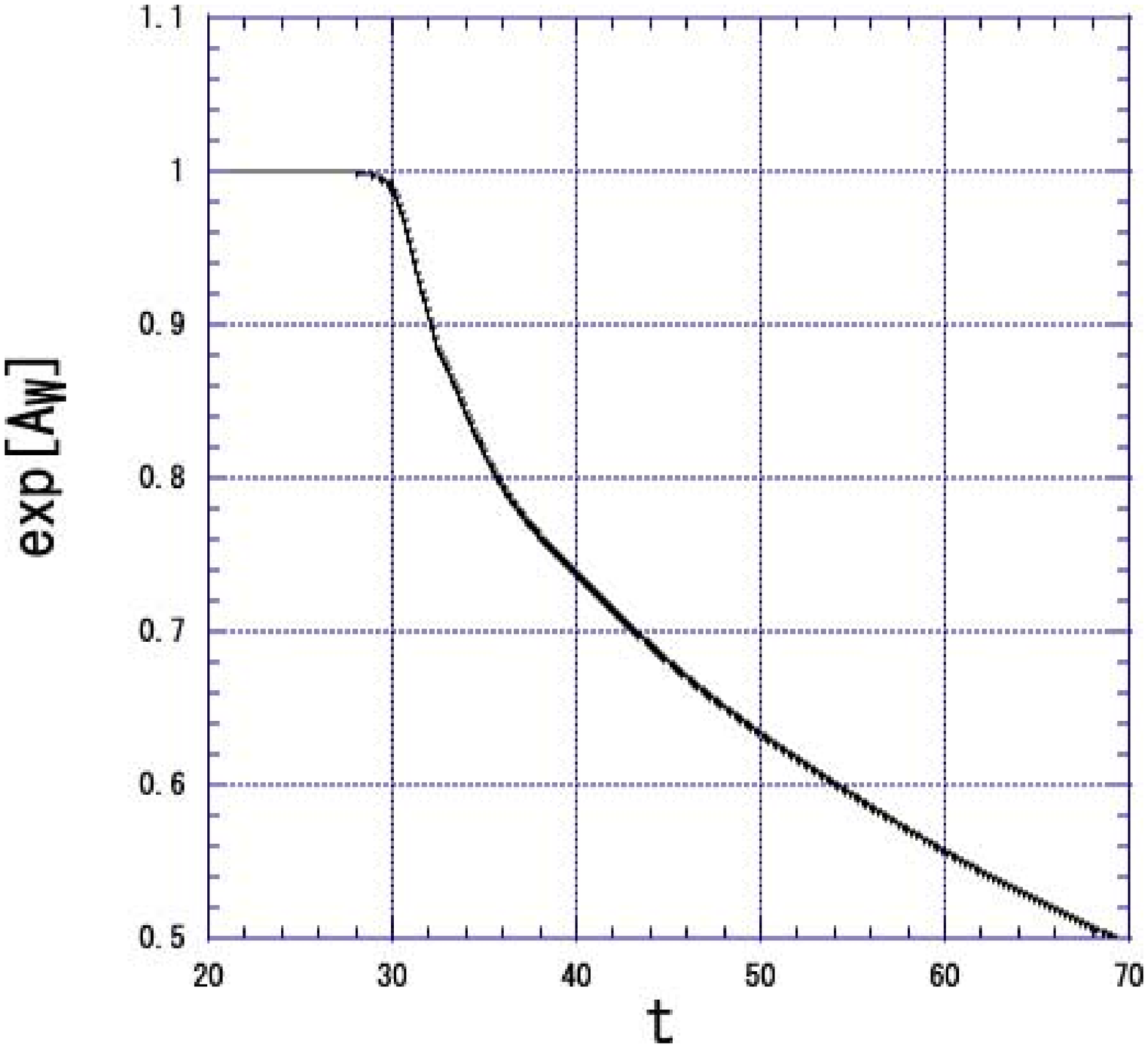} &
\includegraphics[width=4cm]{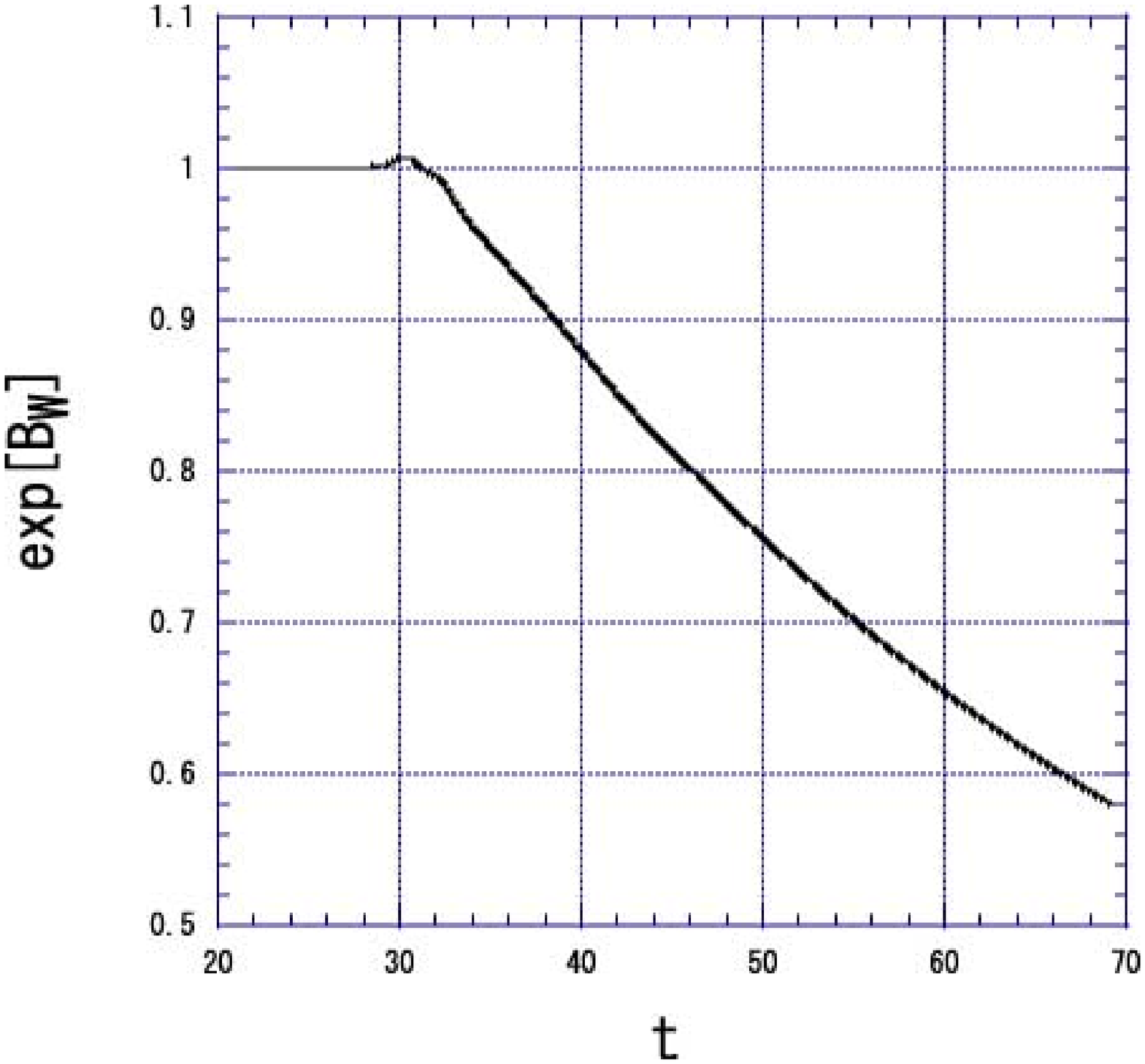}\\
$A_{\rm W}$ & $B_{\rm W}$ \\
\end{tabular}
\caption{Time evolution of the metric $A$ and $B$ 
on the moving wall 
for $\upsilon=0.4,\ d=\sqrt{2},\ \kappa_5=0.15$. 
The value of the metric is given by  
$A_{\rm W}=A(t, z_{\rm W}(t))$, where $z_{\rm W}(t)$ is 
the position of the wall. Both of two quantities decreases with time.}
\label{fig6}
\end{center}
\end{figure}
%%%%%%%%%%%%%%%%%%%%%%%%%%%%%%%%
%%%%%%%%%%%%%%%%%%%%%%%%%%%%%%
\begin{figure}[htbp]
\begin{center}
\begin{tabular}{cc}
\includegraphics[width=4cm]{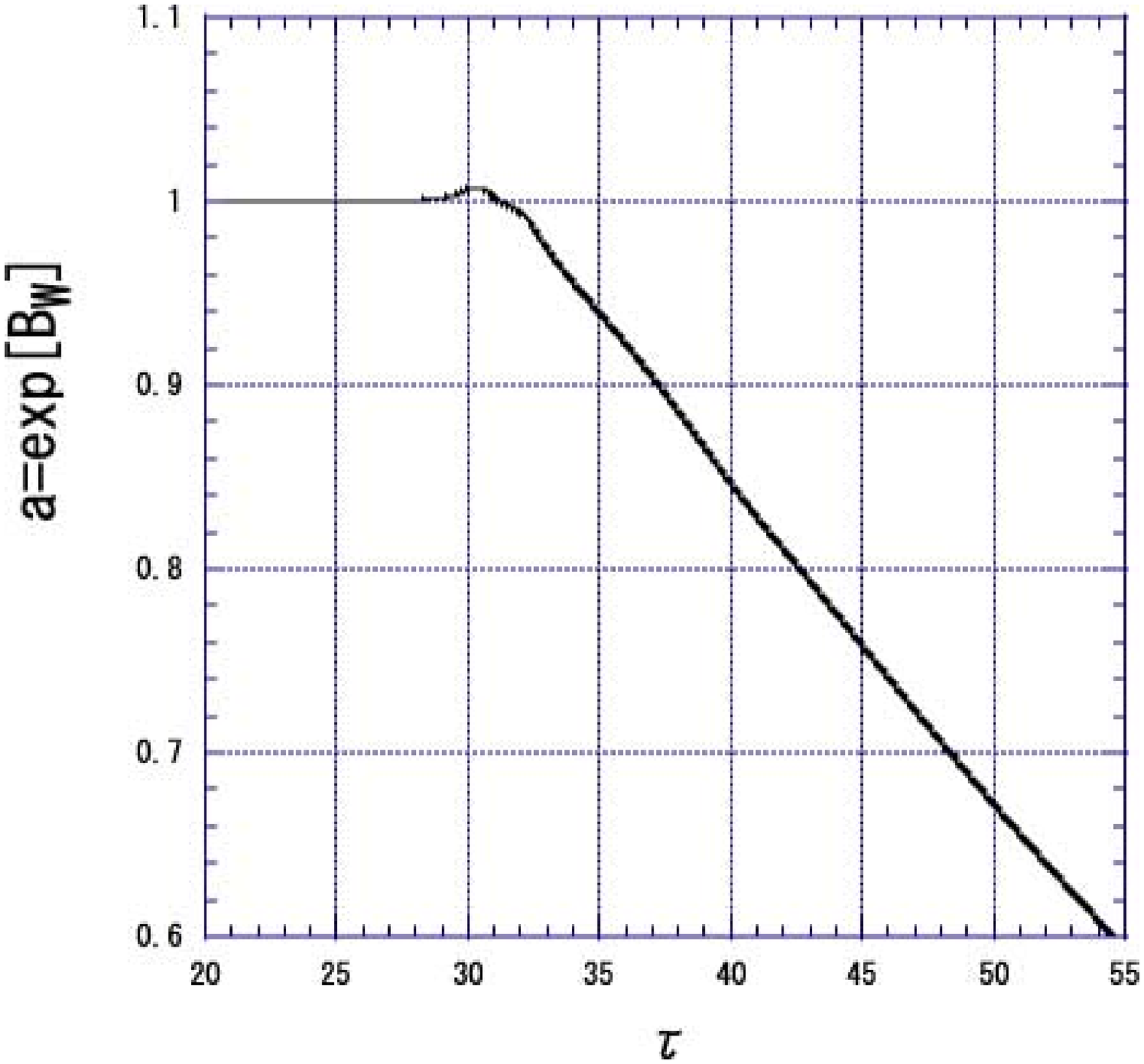} &
\includegraphics[width=4cm]{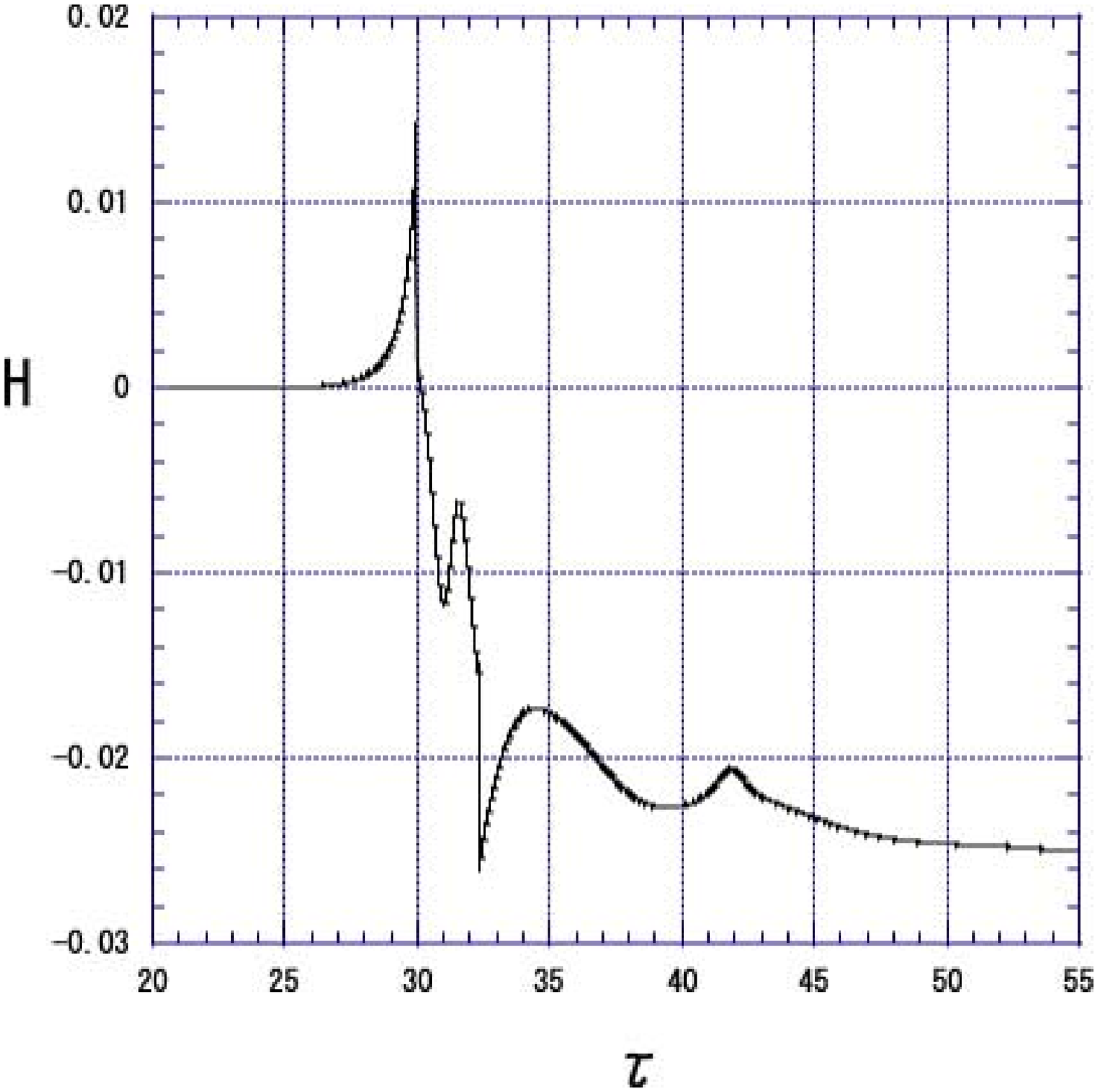}\\
(a) $a$ & (b) $H$ \\
\end{tabular}
\caption{Time evolution of the scale factor $a=e^{B_{\rm W}}$ and 
the Hubble parameter 
$H\equiv \dot{a}(\tau)/a$
for $\upsilon=0.4,\ d=\sqrt{2},\ \kappa_5=0.15$ with respect to 
the proper time $\tau$.}
\label{fig10}
\end{center}
\end{figure}
%%%%%%%%%%%%%%%%%%%%%%%%%%%%%%%%%%%

We evaluate the time evolutions of the metric $A_{\rm W}, B_{\rm W}$ 
on the brane and plot them in Figs. \ref{fig6}. Both of 
two quantities decrease with time except that $B_{\rm W}$ increases slightly 
through the bounce. 

\begin{figure}[htbp]
\begin{center}
\includegraphics[width=8cm]{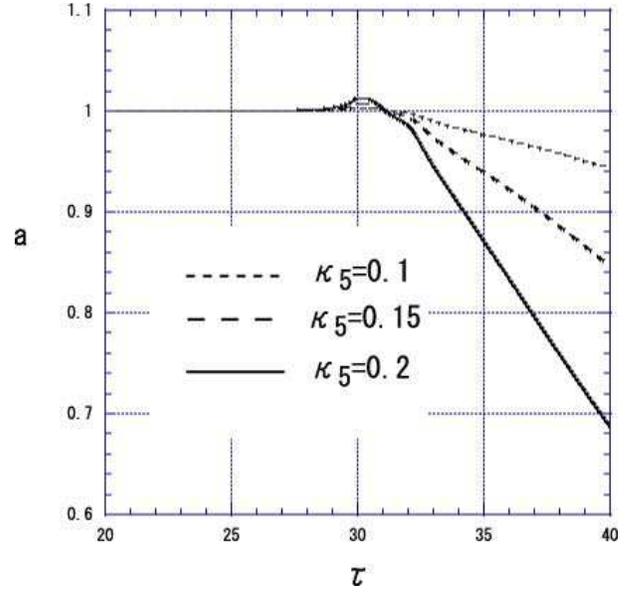}
\caption{Time evolution of the scale factor $a=e^{B_{\rm W}}$ 
for $\upsilon=0.4,\ d=\sqrt{2}$. We set $\kappa_5=0.1, 0.15$
and 0.2. 
As $\kappa_5$
gets larger, the speed of contraction becomes larger.} 
\label{scale_com}
\end{center}
\end{figure}

\begin{figure}[htbp]
\begin{center}
\includegraphics[width=8cm]{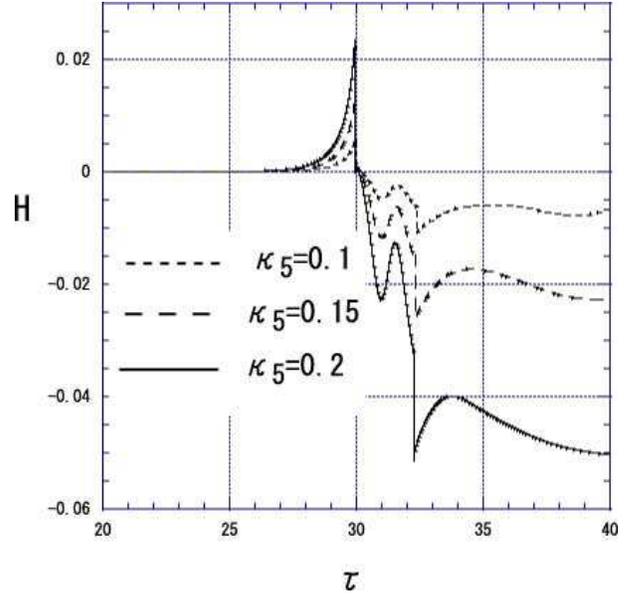}
\caption{Time evolution of the Hubble parameter  $H
\equiv \dot{a}(\tau)/a$ 
for $\upsilon=0.4,\ d=\sqrt{2}$. We set $\kappa_5=0.1, 0.15$
and 0.2. 
As $\kappa_5$
gets larger, the typical time scales of expansion and contraction 
become larger.}
\label{hubble_com}
\end{center}
\end{figure}

 From those two quantities, using Eqs. (\ref{scale factor}) and 
(\ref{hubble}), we evaluate 
a scale factor of our universe 
$a(\tau)=e^{B_{\rm W}}(\tau)$ and the Hubble expansion 
parameter $H \equiv {da\over d\tau}/a$, where 
$\tau$ is the proper time of domain wall defined by Eq. (\ref{proper time}), 
and show them in Figs. \ref{fig10} setting $\upsilon=0.4$.
 From this figure, we find that
 our universe expands slightly before bounce then eventually contracts. 
For each $\kappa_5$, the scale factor $a$ and the Hubble parameter
$H$ are plotted in 
Figs. \ref{scale_com} and \ref{hubble_com}. From these figures, 
we see that
our universe contracts faster as $\kappa_5$ gets larger, i.e.
a negative cosmological constant increases.
%%%%%%%%%%%%%%%%%%%%%%%%%%%%%%
%%%%%%%%%%%%%%%%%%%%%%%%%%%%%%%%%%%
%%%%%%%%%%%%%%%%%%%%%%%%%%%%%%
%%%%%%%%%%%%%%%%%%%%%%%%%%%%%%%%%%%
%%%%%%%%%%%%%%%%%%%%%%%%%%%%%%%%%%%
\begin{figure}[htbp]
\begin{center}
\begin{tabular}{cc}
\includegraphics[width=4cm]{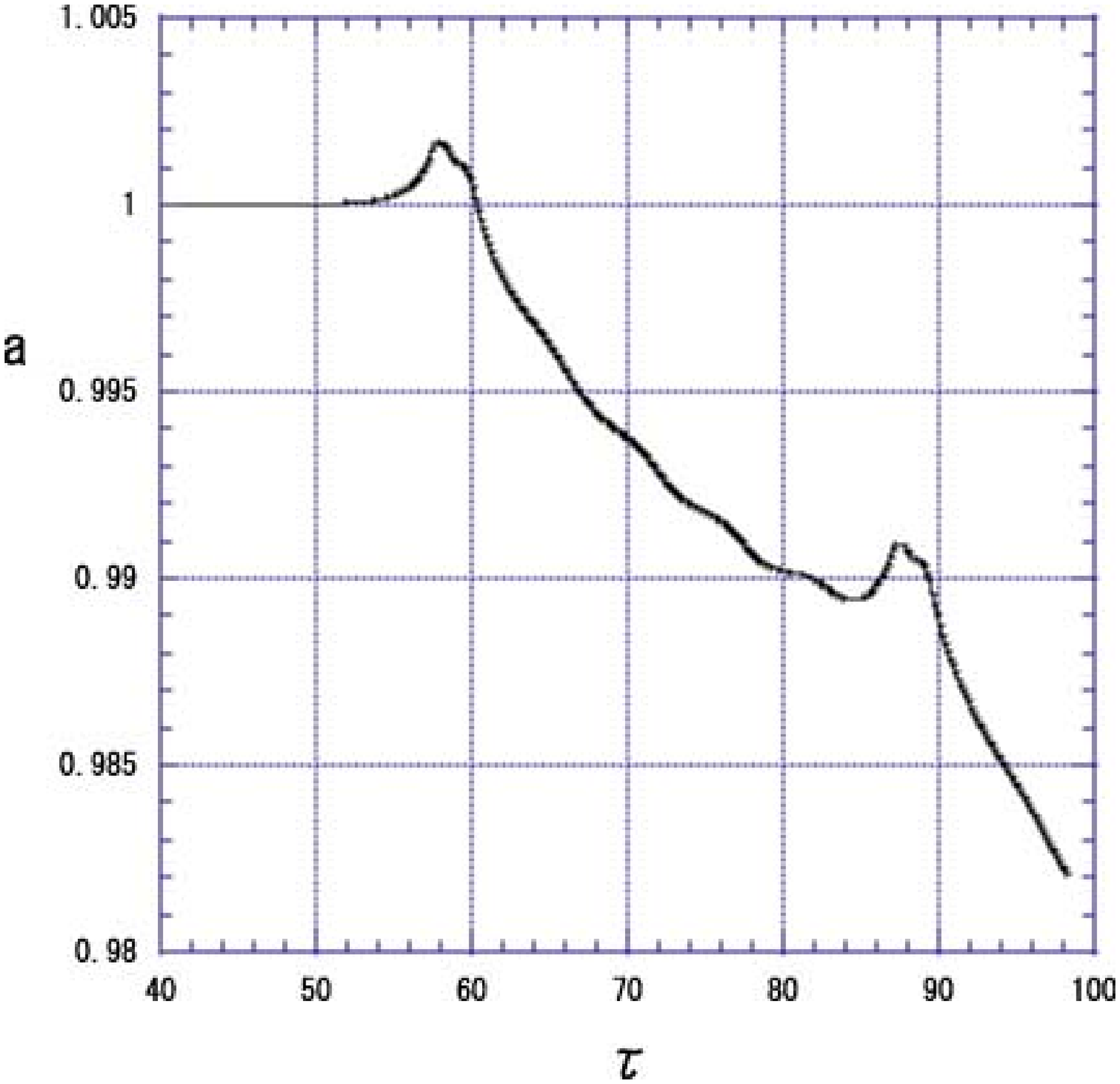} &
\includegraphics[width=4cm]{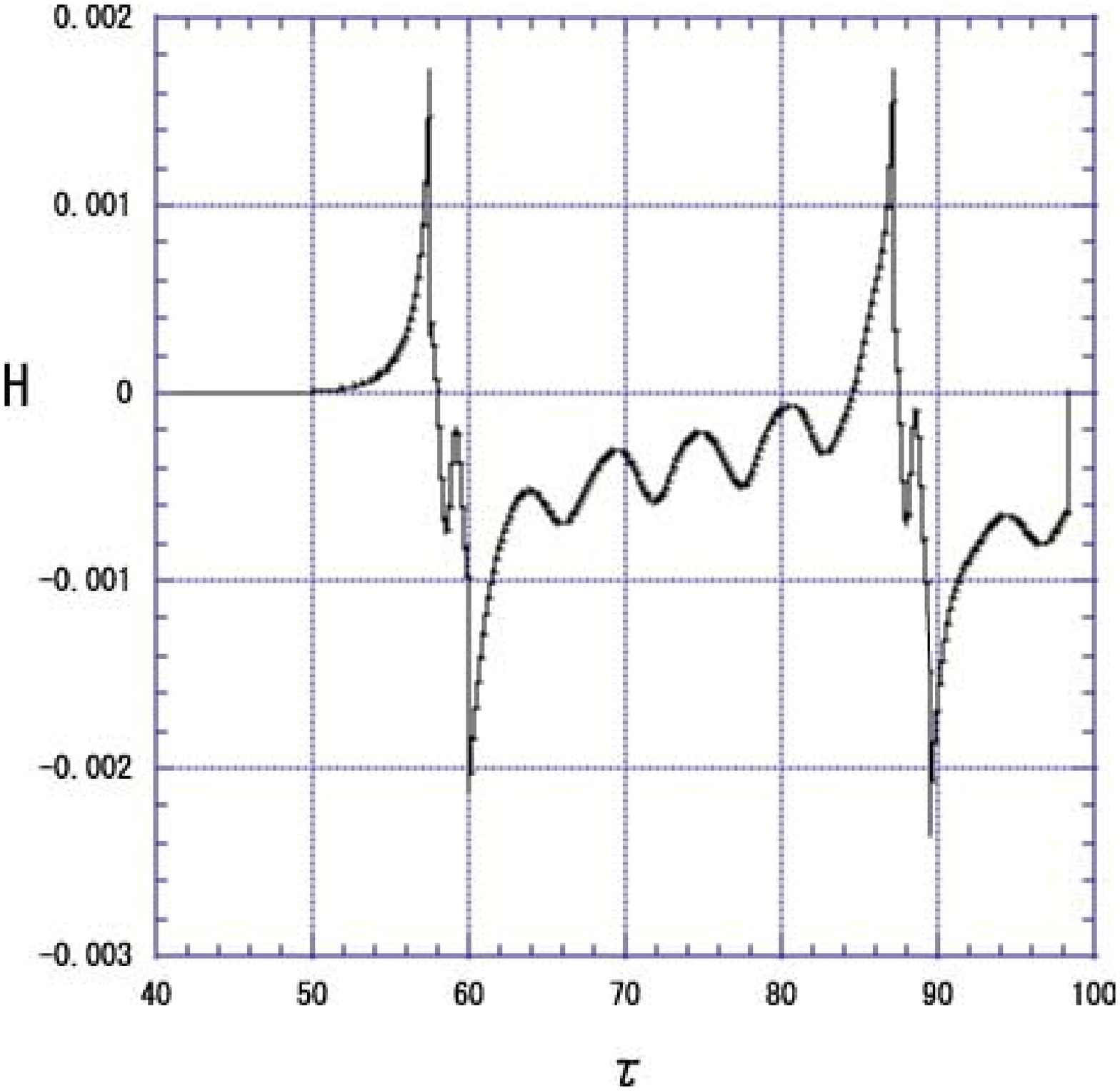}\\
(a) $a$ & (b) $H$ \\
\end{tabular}
\caption{Time evolution of the scalar factor $a$ and the Hubble parameter
$H$
on the moving wall for $\upsilon=0.2,\ d=\sqrt{2}, \ \kappa_5=0.05$. We 
find two contracting phases.}
\label{v=0.2_metric}
\end{center}
\end{figure}
%%%%%%%%%%%%%%%%%%%%%%%%%%%%%%%%
%%%%%%%%%%%%%%%%%%%%%%%%%%%%%%%%%%%
\begin{figure}[htbp]
\begin{center}
\begin{tabular}{cc}
\includegraphics[width=4cm]{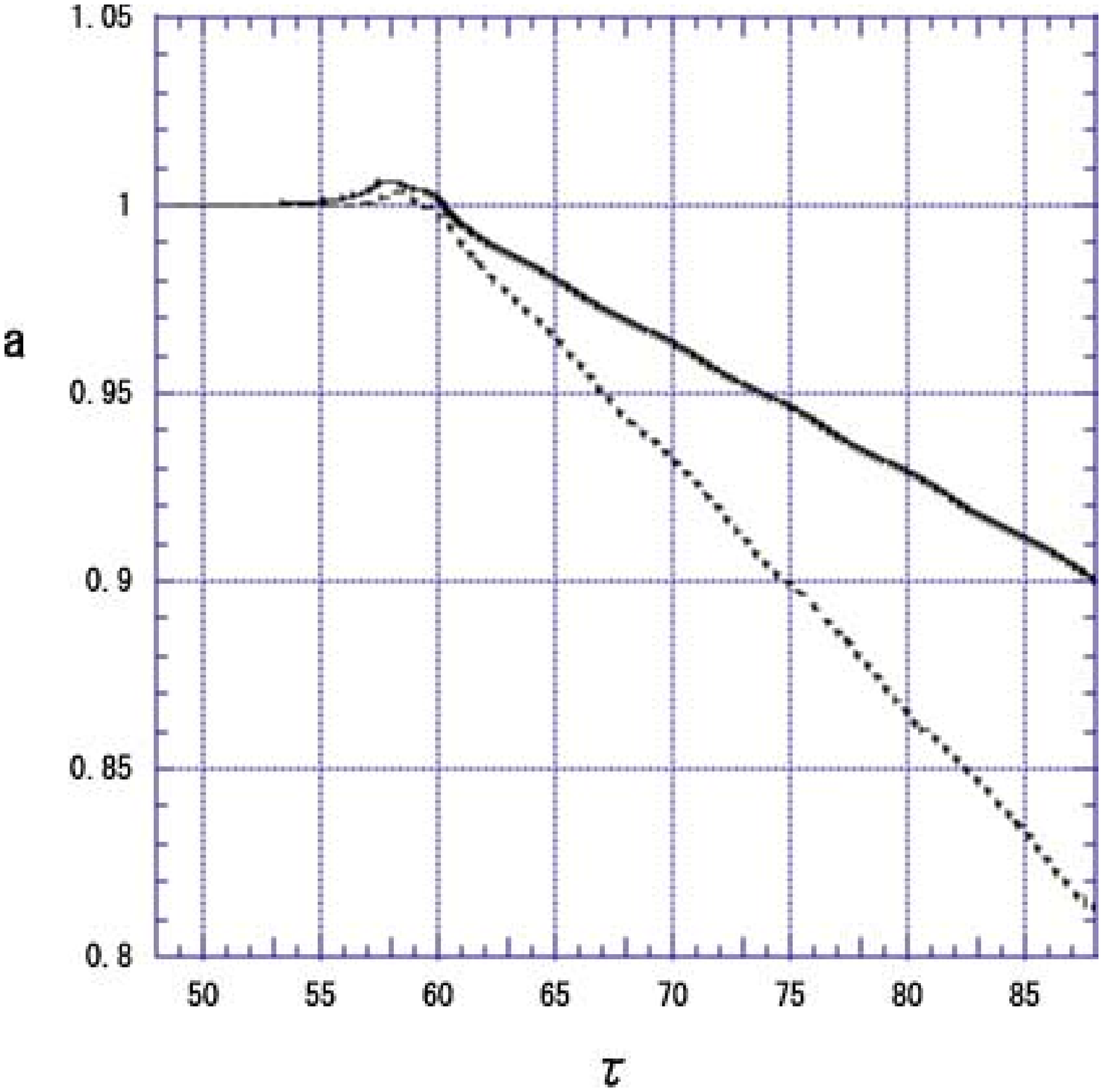} &
\includegraphics[width=4cm]{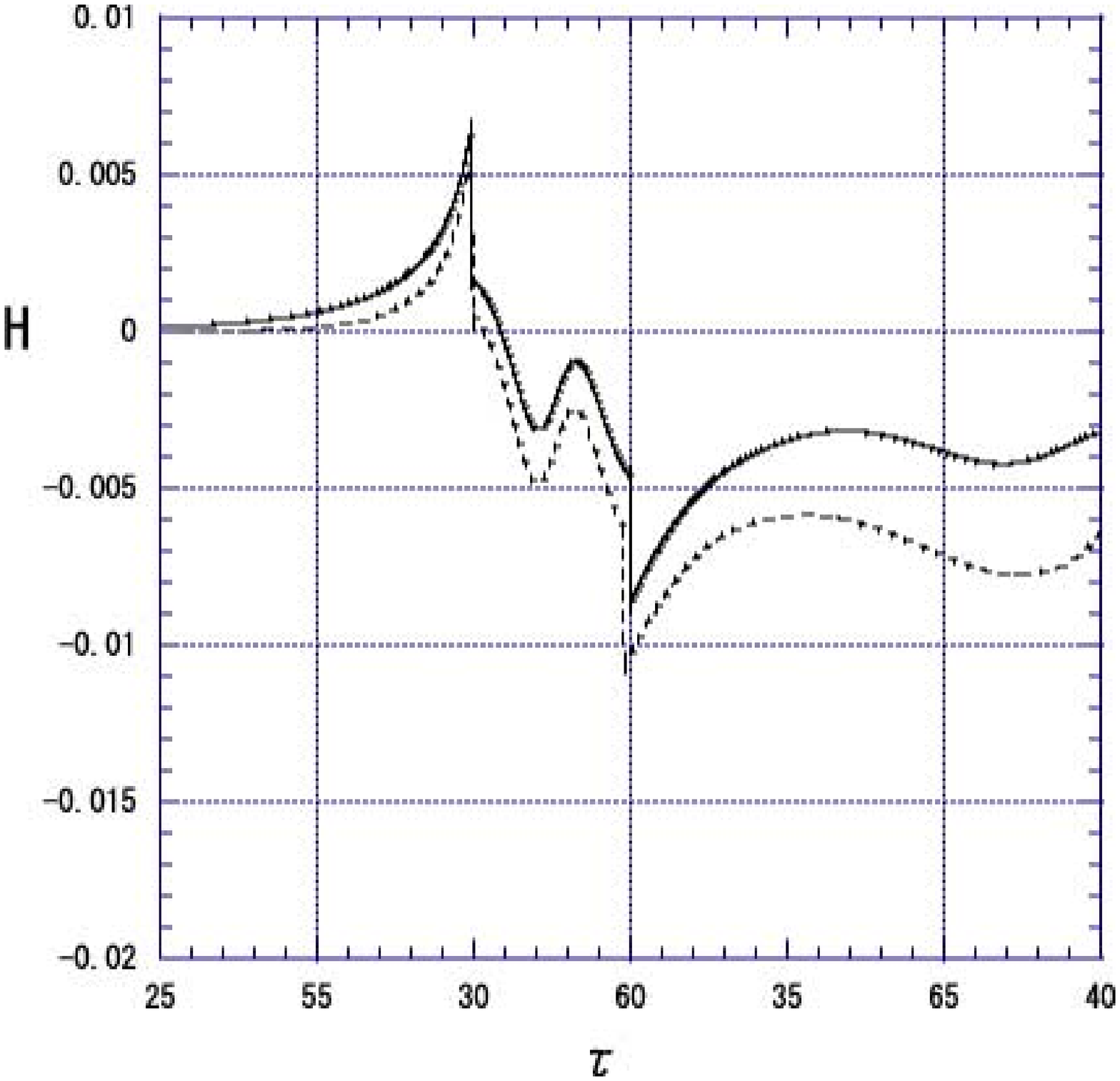}\\
(a) $a$ & (b) $H$ \\
\end{tabular}
\caption{Time evolution of the scalar factor $a$
and the Hubble parameter $H$
on the moving wall for $\upsilon=0.2,\ d=\sqrt{2}, \ \kappa_5=0.1$. The dashed
and dotted lines 
denote the cases of 
$\upsilon=0.2$ and  of $\upsilon=0.4$, respectively. We 
find the universe contracts slower in the case of 
$\upsilon=0.2$ than in the case $\upsilon=0.4$. }
\label{v=0.2_metric2}
\end{center}
\end{figure}
%%%%%%%%%%%%%%%%%%%%%%%%%%%%%%%%

Next we investigate the scale factor $a$ and the Hubble
parameter $H$ for the case of $\upsilon=0.2$. 
Setting $\kappa_5=0.05$, that is the case of 
two-bounce collision, 
$a$ and $H$ are plotted in Fig. \ref{v=0.2_metric}. 
We find two contracting phases, 
which correspond to each bounce at collision.

For $\kappa_5\gsim 0.1$, the bounce  occurs once only because of 
a large negative cosmological constant. 
$a$ and $H$ are plotted 
in Fig. \ref{v=0.2_metric2} for $\kappa_5=0.1$. In this figure, 
we find the universe contracts slower than the case $\upsilon=0.4$. 
Our universe contracts faster as $\upsilon$ gets larger. 

%%%%%%%%%%%%%%%%%%%%%%%%%%%%%%%%%%%
\begin{figure}[htbp]
\begin{center}
\begin{tabular}{cc}
\includegraphics[width=4cm]{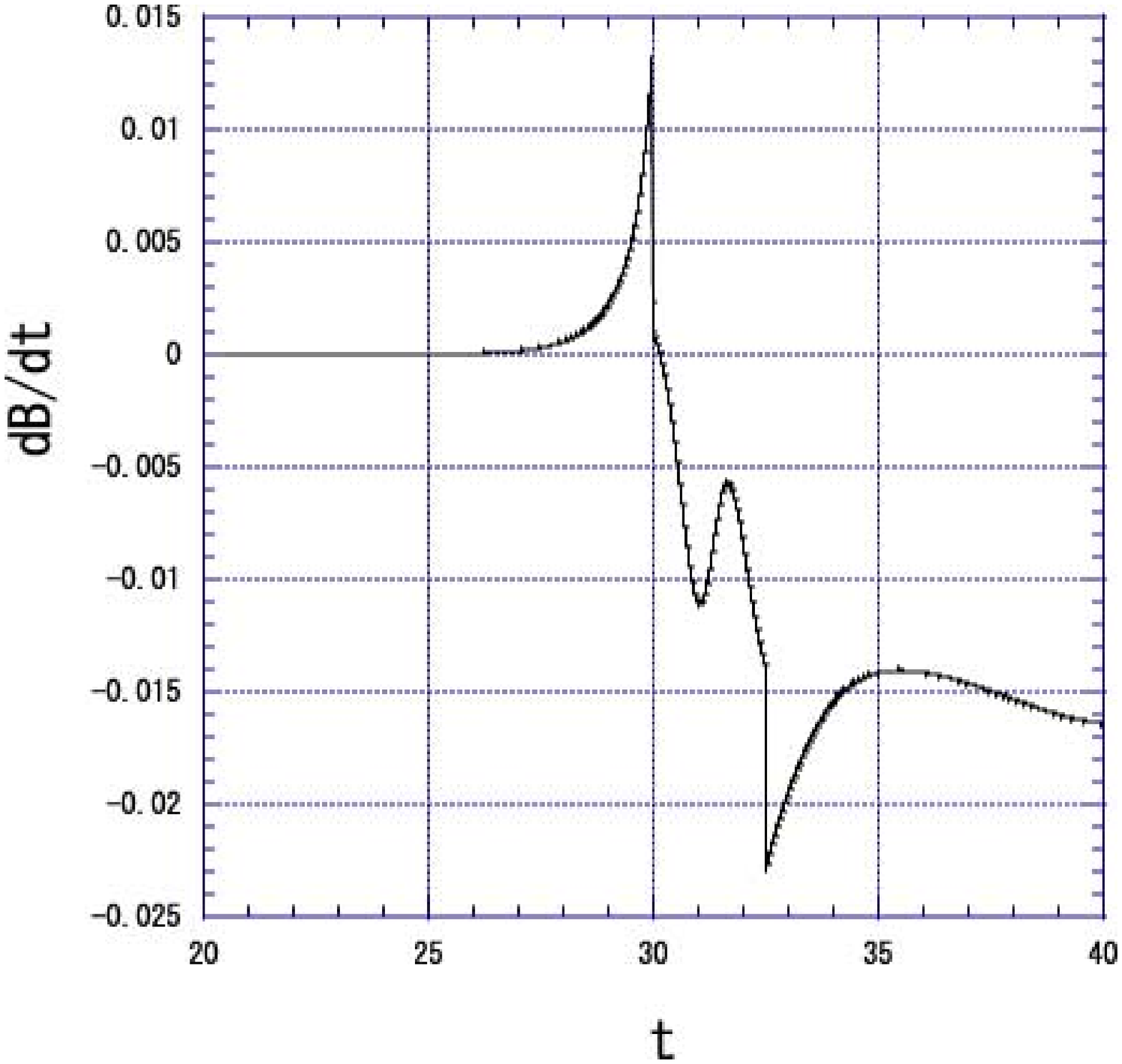} &
\includegraphics[width=4cm]{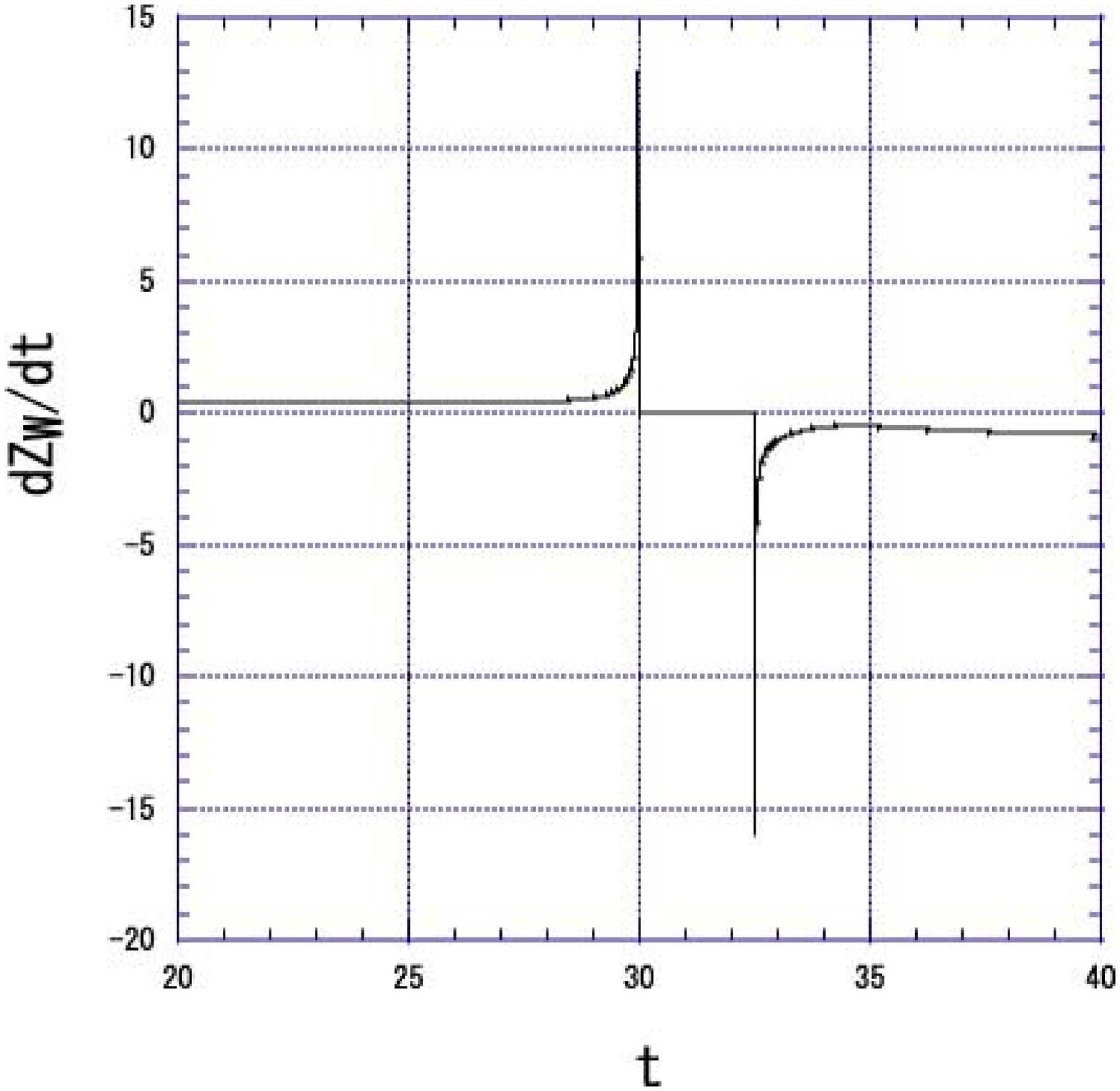}\\
$\dot{B}$ & $\dot{z}_{\rm W}$ \\
\end{tabular}
\caption{Time evolution of $\dot{B}$ and 
$\dot{z}_{\rm W}$ for
$\upsilon=0.4,\ d=\sqrt{2}, \ \kappa_5=0.15$. We find 
two discontinuous stages at $t\sim 30$ and $t\sim 32$. 
}
\label{discontinuous}
\end{center}
\end{figure}
%%%%%%%%%%%%%%%%%%%%%%%%%%%%%%%%

Finally, we find  in Fig. \ref{hubble_com}
that there are two discontinuous stages in the evolution of 
the Hubble parameter at $t\sim 30$ and $\sim 32$. 
As we will see, we can conclude that
these discontinuities appear just because of 
ambiguity of the definition of a wall-position, 
$z=z_{\rm W}$. The Hubble parameter  mainly depends on
the derivative of the metric 
with respect to $t$, $\dot{B}$.
This quantity in fact has 
two discontinuous stages as 
seen in Fig. \ref{discontinuous}. 
In this paper, we define the position of 
wall by one where the energy density of scalar field gets maximum. 
However, 
this position does not move continuously through a bounce. Actually, 
we show $\dot{z}_{\rm W}$ has also two discontinuous stages 
as shown in Fig. \ref{discontinuous}, where the 
speed of a wall apparently exceeds the speed of light. 
Namely near the bounce, the definition of 
wall-position is not well-defined.
This is because there is no wall configuration during the collision. 
Hence these discontinuities of the Hubble parameter
seem to be apparent.
We should look at the global time evolution.

%%%%%%%%%%%%%%%%%%%%%%%%%%%%%%%%%
%%%%%%%%%%%%%%%%%%%%%%%%%%%%%%%%%
\section{summary and discussion}
%%%%%%%%%%%%%%%%%%%%%%%%%%%%%%%%%
%%%%%%%%%%%%%%%%%%%%%%%%%%%%%%%%%
We have studied collision of two domain walls 
in 5D asymptotically Anti de Sitter spacetime. This 
may provide the reheating mechanism of an ekpyrotic 
(or cyclic) brane universe, in which two BPS branes collide and evolve 
into a hot big bang universe. We evaluate the values of both a scalar field 
corresponding to a domain wall and 
metric on the moving wall for different value of the warp factor $k$
 which is related to a gravitational effect $\kappa_5$. 
We analyze two typical incident velocities, i.e.  $\upsilon=0.4$, and $\upsilon=0.2$, which correspond  to 
one-bounce and two-bounce solutions
in the Minkowski spacetime, respectively. 

For the case of $\upsilon=0.4$, the global feature of
collision does not change so much for 
different values of $\kappa_5$, 
but the behaviour of oscillation after the collision 
is different for each $\kappa_5$. 
For small value of $\kappa_5\lsim 0.01$, the oscillation 
is the same as Minkowski case, but for $\kappa_5\gsim0.05$, it
becomes an overstable oscillation. 
So its period and amplitude get larger as $\kappa_5$ 
increases. In the cause of this unstable oscillation, the singularity appears
after collision. This singularity 
is very similar to that found in Khan and Penrose \cite{Khan_Penrose},
in which they discuss  collision of plane waves and formation
 of a singularity. 
Hence the appearance of singularity in the present
model could be understandable because we take  into account a
gravitational effect in collision of two domain walls.

In the time evolution of our universe, 
we find that the universe first expands a little just 
before collision and then 
contracts just after collision.
This result is consistent with \cite{LMW}.
We cannot explain our hot big bang universe as it is. 
It is also found that 
the speed of expansion and contraction gets faster as 
$\kappa_5$ increases. 

For the second case, i.e.,  $\upsilon=0.2$, we show the bounce does not 
occur twice for larger value of $\kappa_5$ ($\kappa_5\gsim 0.1$) 
corresponding to the unstable oscillation. 

We shall discuss about the value of a warp factor $k$.  
We consider a curvature length $l=1/k$ written in the following form 
\begin{equation}
l=1.97\times 10^{-17}\Bigl(\frac{10^{-2}}{k}\Bigr)
 \Bigl(\frac{{\rm TeV}}{m_{\Phi}}\Bigr)~[{\rm m}] \,,
\end{equation}
where $m_{\Phi}$ is a mass scale of a domain wall. 
Here we set a value of a warp factor $k$ in the region 
$0.01\lsim k \lsim 0.25$.
On the other hand, we also know 
from the experimental data of testing a gravitational 
inverse-square law that the curvature length must be smaller 
than $0.1$ mm
\cite{0.1mm}. 
From this constraint equation, we obtain 
\begin{equation}
k> 1.97\times 10^{-15}\Bigl(\frac{{\rm TeV}}{m_{\Phi}}\Bigr)\,.
\end{equation}
So the values of $k$ used
in our simulation satisfy this constraint.

\acknowledgments

We would like to thank H. Kudoh and S. Mizuno for useful discussions.
This work was partially supported by the Grant-in-Aid for Scientific Research
Fund of the JSPS (No. 17540268) and another for the 
Japan-U.K. Research Cooperative Program,
and by the Waseda University Grants for Special Research Projects and 
 for The 21st Century
COE Program (Holistic Research and Education Center for Physics
Self-organization Systems) at Waseda University.

%%%%%%%%%%%%%%%%%%%%%%%%%%%%%%%%%
%%%%%%%%%%%%%%%%%%%%%%%%%%%%%%%%%
\appendix
%%%%%%%%%%%%%%%%%%%%%%%%%%%%%%%%%
%%%%%%%%%%%%%%%%%%%%%%%%%%%%%%%%%%%%%%%%%%%%
\section{Perturbations of a domain wall\label{appen:perturb}}
%%%%%%%%%%%%%%%%%%%%%%%%%%%%%%%%%%%%%%%
In Fig. \ref{fig5}, we find oscillations after collision and 
those amplitudes and periods 
increase as $\kappa_5$ gets larger. To understand this feature, 
we analyze perturbations
around a static domain wall solution
 in this appendix. 

We use the coordinate $y$, by which 
a static domain wall solution is 
given by analytically ($\Phi_K(y), A_K(y)$),
which are given by Eqs. (\ref{kink}) and (\ref{asym_AdS_metric}).
We perturb the 
basic equations (\ref{Dynamical equation}) and  
(\ref{Constraint equation}) 
by setting $A=A_K(y)+a(t,y)$, $B=A_K(y)+b(t,y)$, and 
$\Phi=\Phi_K(y)+\phi(t,y)$.
We find
two sets of perturbation equations:\\
(1) dynamical equations
\begin{eqnarray}
e^{-2A_K} \ddot{a} =&&
{\partial^2 a \over \partial y^2}+{dA_K \over dy }{\partial  
a \over \partial y }
-6{dA_K \over dy } {\partial  b \over \partial y } \nonumber\\
&&+\kappa_5^2\left(
2{d \Phi_K \over d y }{\partial\phi\over \partial y}
-{2\over 3}V\big|_K a-{1\over 3}{dV\over d\Phi}\Big{|}_K\phi
\right)\,,\label{dynamical:a}
\\
e^{-2A_K}\ddot{b}=&&
{\partial^2 b \over \partial y^2}+7{dA_K \over dy }{\partial  
b \over \partial y }
+
{2\over 3}\kappa_5^2\left(2V\big|_K a+
{dV\over d\Phi}\Big{|}_K\phi\right)
\,,
\label{dynamical:b}
\nonumber\\
\\
e^{-2A_K} \ddot{\phi} =&&
{\partial^2 \phi \over \partial y^2}+4{dA_K \over dy }{\partial  
\phi \over \partial y }
+3{d \Phi_K \over d y }{\partial  b \over \partial y }\nonumber\\
&&-{1\over 2}\left(2{dV\over d\Phi}\Big{|}_K a+
{d^2V\over d\Phi^2}\Big{|}_K\phi\right)
\label{dynamical:phi}
\,,
\end{eqnarray}
(2) constraint equations
\begin{eqnarray}
&&
{\partial\dot{b}\over \partial y}-{dA_K\over d y}\dot{a}
=-{2\over 3}\kappa_5^2{d\Phi_K\over d y}\dot{\phi}
\,,
\label{const1}
\\
&&
{\partial^2 b\over \partial y^2}
+4{dA_K \over dy }{\partial b\over \partial y}
-{dA_K \over d y }{\partial  
a \over \partial y }\nonumber\\
&&+{2\over 3}\kappa_5^2
\left(V\big|_K a+
\frac{1}{2}{dV\over d\Phi}\Big{|}_K\phi+{d\Phi_K \over dy} 
{d\phi \over dy} \right)
=0\label{const2}
\end{eqnarray}
In order to find the eigenvalue and eigen functions,
we set $a=\tilde{a}(y)e^{i\omega t}$, $b=\tilde{b}(y)e^{i\omega t}$, and 
$\phi=\tilde{\phi}(y)e^{i\omega t}$.
Then the constraint equation (\ref{const1}) is reduced to be
\begin{eqnarray}
{d \tilde{b}\over d y}-{dA_K\over d y}\tilde{a}
=-{2\over 3}\kappa_5^2{d\Phi_K\over d y}\tilde{\phi}
\label{const11}
\end{eqnarray}
Inserting Eq. (\ref{const11}) into another constraint (\ref{const2}) and 
using the equations for a background solution,
we find that the constraint equation (\ref{const2}) 
turns out to be trivial. 
So we have only one constraint equation (\ref{const11}). 

Eliminating ${d\til{b}/dy }$ in 
(\ref{dynamical:a}) and (\ref{dynamical:phi})
by use of Eq. (\ref{const11}),  we obtain two coupled
perturbation equations  in terms of $\til{a}, \til{\phi}$ as 
\begin{eqnarray}
&&
{d^2 \tilde{a} \over d y^2}=
\left[6\left({dA_K \over dy }\right)^2
+{2\over3}\kappa_5^2V\big|_K-e^{-2A_K}\omega^2
\right]\til{a} 
-{dA_K \over dy }{d\tilde{a} \over d y }\nonumber\\
&&-\kappa_5^2\left(4{dA_K\over dy}{d\Phi_K\over dy}-\frac{1}{3}{dV\over d\Phi}
\Big{|}_K\right)\til{\phi}-2\kappa_5^2{d\Phi_K\over dy}{d\til{\phi}\over dy}
\,,
\label{eigen:a}
\\[.5em]
%\end{eqnarray}
%\begin{eqnarray}
&&
{d^2 \tilde{\phi} \over d y^2}=
\left({dV\over d\Phi}\Big{|}_K-3{d \Phi_K \over d y }
{dA_K\over dy}\right)\til{a}-4{dA_K \over dy }{d\tilde{\phi} \over d y }
\nonumber\\
&&+\left[2\kappa_5^2\left({d\Phi_K\over dy}\right)^2
+\frac{1}{2}{d^2V\over d\Phi^2}\Big{|}_K-e^{-2A_K}\omega^2\right]
\til{\phi}
\,.
\label{eigen:phi}
\end{eqnarray}
Eq. (\ref{dynamical:b}) 
is guaranteed by the other two dynamical equations 
and constraint equations.
Eqs. (\ref{eigen:a})
and (\ref{eigen:phi}) have the asymptotic
forms as $y\to \infty$ 
as 
\begin{eqnarray}
\til{a}&=&e^{\pm \sqrt{A_1} y}\,,\label{asym:tila}\\
\til{\phi}&=&e^{\pm \sqrt{A_2} y}\,,\label{asym:tilp}
\end{eqnarray}
where 
\begin{eqnarray}
&&A_1=-e^{-2A_\infty}\omega^2\,,\\
&&A_2={1\over2}\frac{d^2V}{d\Phi^2}\Big|_{\Phi=1}
-e^{-2A_\infty}\omega^2 \,.
\end{eqnarray}
 Here we choose both negative signs in Eqs. (\ref{asym:tila})
and (\ref{asym:tilp})
 because 
negative signs correspond to out-going wave modes.

We solve numerically these equations (\ref{eigen:a}, \ref{eigen:phi}) 
connecting the above asymptotic solutions and find the complex 
eigen frequency $\omega$. 
For $\kappa_5=0.1$, we obtain a stable mode as 
$\omega_{\rm s}
=1.23+ 1.07\times 10^{-3} i$ and unstable mode as 
$\omega_{\rm u}
=0.644 -2.90\times 10^{-2} i $. Compared this unstable mode with 
the value obtained from the oscillations after collision found in 
Fig. \ref{fig5} 
(Notice that we use a proper time $\tau$ in Fig. \ref{fig5}.
Then it should be evaluated in the physical time $t$). 
We obtain $0.772\approx
1.2 \times \Re [\omega_{\rm u}]$ for the real part of the frequency 
and $-0.029\approx
1.0 \times \Im [\omega_{\rm u}]$ for the imaginary part.
 So we may conclude that
the overstable oscillations 
after the collision of two domain walls found in Fig. \ref{fig5}
are explained by
the unstable mode around a static domain wall
solution. 

%%%%%%%%%%%%%%%%%%%%%%%%%%%%%%%%%

\end{document}